\documentclass[reprint,aps,prb,amsmath,amssymb,twocolumn,showkeys, superscriptaddress,floatfix,showkeys,twoside]{revtex4-2}
\usepackage{graphicx}
\usepackage{dcolumn}
\usepackage{bm}
\usepackage[colorlinks = true, allcolors = blue]{hyperref}
\usepackage{verbatim}
\usepackage{soul}
\usepackage{float}
\usepackage{xcolor}
\usepackage{multirow}
\usepackage{fancyhdr}
\begin{document}

\title{Retrieval of material properties of monolayer transition-metal dichalcogenides\\from magnetoexciton energy spectra}

\author{Duy-Nhat Ly}
\email{nhatld@hcmue.edu.vn}
\affiliation{%
 Computational Physics Key Laboratory K002, Department of Physics, Ho Chi Minh City University of Education, Ho Chi Minh City 72759, Vietnam
}%
\thanks{D.-N. Ly and D.-N. Le contributed equally to this work.}

\author{Dai-Nam Le}
\email{dainamle@usf.edu}
\affiliation{%
 Department of Physics, University of South Florida, Tampa, FL 33620, United States of America
}%
\thanks{D.-N. Ly and D.-N. Le contributed equally to this work.}

\author{Duy-Anh P. Nguyen}
\affiliation{The Institute of Applied Technology, Thu Dau Mot University, Thu Dau Mot City, Binh Duong Province, Vietnam
}%

\author{Ngoc-Tram D. Hoang}
\affiliation{%
 Computational Physics Key Laboratory K002, Department of Physics, Ho Chi Minh City University of Education, Ho Chi Minh City 72759, Vietnam
}%

\author{Ngoc-Hung Phan}
\affiliation{%
 Computational Physics Key Laboratory K002, Department of Physics, Ho Chi Minh City University of Education, Ho Chi Minh City 72759, Vietnam
}%

\author{Hoang-Minh L. Nguyen}
\affiliation{%
 Computational Physics Key Laboratory K002, Department of Physics, Ho Chi Minh City University of Education, Ho Chi Minh City 72759, Vietnam
}%

\author{Van-Hoang Le}
\email{hoanglv@hcmue.edu.vn}
\affiliation{%
 Computational Physics Key Laboratory K002, Department of Physics, Ho Chi Minh City University of Education, Ho Chi Minh City 72759, Vietnam
}%

\date{\today}

\begin{abstract}
Reduced exciton mass, polarizability, and dielectric constant of the surrounding medium are essential properties for semiconducting materials, and they have been extracted recently from the magnetoexciton energies. However, the acceptable accuracy of the suggested method requires very high magnetic intensity. Therefore, in the present paper, we propose an alternative method of extracting these material properties from recently available experimental magnetoexciton s-state energies in monolayer transition-metal dichalcogenides (TMDCs). The method is based on the high sensitivity of exciton energies to the material parameters in the Rytova-Keldysh model. It allows us to vary the considered material parameters to get the best fit of the theoretical calculation to the experimental exciton energies for the $1s$, $2s$, and $3s$ states. This procedure gives values of the exciton reduced mass and $2D$ polarizability. Then, the experimental magnetoexciton spectra compared to the theoretical calculation also determine the average dielectric constant. Concrete applications are presented only for monolayers WSe$_2$ and WS$_2$ from the recently available experimental data; however, the presented approach is universal and can be applied to other monolayer TMDCs.
The mentioned fitting procedure requires a fast and effective method of solving the Schr\"{o}dinger equation of an exciton in monolayer TMDCs with a magnetic field. Therefore, we also develop such a method in this paper for highly accurate magnetoexciton energies.
\end{abstract}


\maketitle

\section{\label{intro}Introduction}
Two-dimensional van der Waals semiconductors such as transition-metal dichalcogenides (TMDCs) unlock a big door to technological applications such as making ultra-thin computing devices based on their reduced dimensionality, magnetism, (opto-)spintronics, valleytronics or magneto-optics properties \cite{Geim2013, Arora2021, Woods2022, Phan2023}. Especially magnetoexcitons in these materials provide a great potential to make light-control magnetic devices because of their thermal stability as well as their high binding energies. Hence, accurate determination of intrinsic optoelectronic quantities of these monolayer TMDCs, such as their exciton reduced mass, two-dimensional ($2D$) static polarizability, or the dielectric constant of the surrounding medium, is obvious and crucial for future development of designing van-der-Waals-heterostructure-based devices.

There are several methods to determine the exciton reduced mass of monolayer TMDCs. For example, angle-resolved photoemission spectroscopy (ARPES) can experimentally detect energy versus momentum maps and extract effective electron and hole masses \cite{Basov2014, Bussolotti2021, Lee2021, Lin2022}. However, they are expensive and not easy-to-do methods. On the other hand, theoretical studies suggest more effective and accurate ways to determine exciton reduced mass. One of the first methods is estimation from the band structure of \textit{ab initio} calculations, such as density functional theory (DFT) \cite{xiao2012, berkelbach2013, Korm2015}. In recent studies \cite{PRL2018, NAT2019,Liu2019}, optical spectroscopy of magnetoexcitons in monolayer TMDCs has revealed an exciton reduced mass. However, this method utilizes the diamagnetic shift for extraction; thus, it requires a high magnetic intensity for the Landau levels to describe the energy spectra. Based on our estimation, the magnetic fields of 65 and 91 Tesla used in these works must be higher to get an acceptable accuracy, although they have already reached the laboratory limit.

In works \cite{NAT2019,Liu2019}, besides the exciton reduced mass obtained from the experimental diamagnetic shift, other parameters such as the screening length (related to the $2D$ polarizability) and the dielectric constant of the surrounding medium are determined by comparing the experimental data for magnetoexciton energies to the theoretical calculation. Actually, the idea of comparing experimental data with theoretically calculated exciton energies to get the material properties of monolayer TMDCs was suggested early in references \cite{chernikov2014,PhysE}. Especially the study \cite{PhysE} showed that the exciton reduced mass could be extracted from the exciton energies without a magnetic field by the fitting procedure. Therefore, in the present work, we will apply this fitting scheme to the experimental data in \cite{PRL2018, NAT2019} as an alternative method of extracting exciton reduced mass and $2D$ polarizability of monolayer TMDCs. The data with the magnetic field are then used for determining the dielectric constant. The extracted material properties are then compared with data of other works \cite{Plechinger-2016,Stier2016-nat,Stier2016-nano,Zipfel2018,PRL2018,NAT2019,Chen2019-nano,Liu2019}.  

The retrieval method mentioned above requires a combination of highly accurate theoretical calculations of energy spectra and precise experimental measurements of optical spectroscopy of excitons to achieve reliable results. While the experimental data provided in \cite{PRL2018, NAT2019,Liu2019} are the most accurate measurement recently, theoretical energy spectra of the magnetoexciton are nothing but solutions of the Schr\"{o}dinger equation describing a two-dimensional pair of electron and hole that interacts via Rytova-Keldysh potential \cite{Rytova1967, keldysh1979, haramura1988, cudazzo2011} because of the screening effect arising from their reduced dimensionality \cite{berkelbach2013, chernikov2014}. In the case of zero-field, these solutions can be obtained by the variational calculations or semiempirical formula \cite{Molas2019, Hieu2022} with precision enough for analyzing experimental results. However, when a magnetic field or more accurate solutions are needed, we must use a much faster and more precise method. Fortunately, in Ref.~\cite{PhysE}, we have provided exact numerical solutions for some $s$-states of the exciton with and without a uniform perpendicular magnetic field with a precision of up to 20 decimal places by using the so-called Feranchuk-Komarov (FK) operator method \cite{Feranchuk1982, Hoangbook2015}.  In the present study, we even improve this method more advanced by calculating the matrix elements for the Rytova-Keldysh potential using its new integral form that significantly reduces the computational resources compared with the previous version. 

\begin{figure}[htbp]
\begin{center}
\includegraphics[width = 0.9 \columnwidth]{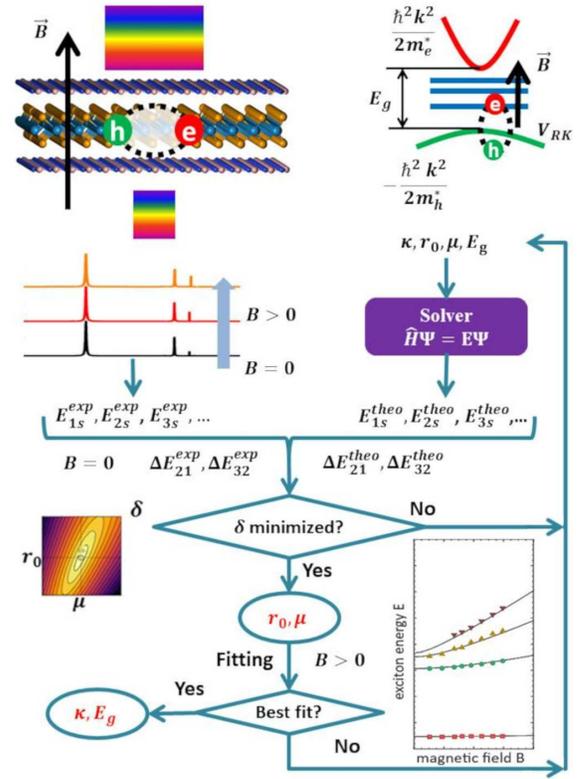}
\caption{Schematic flowchart of extracting the exciton reduced mass, screening length (related to $2D$ static polarizability), dielectric constant and free-particle bandgap of monolayer TMDCs by the fitting scheme for magnetooptical absorption spectra and theoretical solutions of effective Schr\"odinger equation of magnetoexciton.}
\label{fig1}
\end{center}
\end{figure}

Furthermore, examining the sensitivity of magnetoexciton energy on the material parameters allows us to establish an efficient fitting scheme from which we can accurately extract exciton reduced mass, screening length related to $2D$ static polarizability, and dielectric constant from experimental data of optical peaks associated with exciton s-states. We also extract the free-particle bandgap from the experimental exciton energy of the 1s state by comparing it with the calculated one. Hence, a tool with universal data can be developed to retrieve these material properties for any monolayer TMDCs with different substrates. A schematic flowchart is given in 
Fig.~\ref{fig1} to describe our object of study and the method of retrieving the material parameters of monolayer TMDCs.

The rest of this paper is as follows. Section \ref{Sec1} introduces the FK operator method of solving the  Schr\"{o}dinger equation with Rytova-Keldysh potential. Section \ref{Sec2} examines the sensitivity of exciton energy when varying the exciton reduced mass, screening length, and the dielectric constant and then proposes a fitting scheme to retrieve these parameters from experimental data for monolayers WSe$_2$ and WS$_2$. In this section, the 1s exciton energy is also used for determining the free-particle bandgap. Finally, Sec. \ref{conc} includes our conclusions.  

\section{\label{Sec1}Exact numerical solutions for an magnetoexciton in monolayer TMDC}

\textit{Schr{\"o}dinger equation --} 
For a two-dimensional system of one electron and one hole interacting by the potential $\hat V_{h-e}(r)$ in the magnetic field $B {\mathbf e}_z$ perpendicular to the monolayer plane ($x, y$), the center of mass (c.m.) motion can be separated to get the Hamiltonian for the relative motion of the electron and hole as
\begin{equation}
\hat H =\frac{{\hat p}^2}{2 \mu} + \frac{1-\rho}{1+\rho}\frac{eB}{2\mu}{\hat l}_z + \frac{e^2 B^2}{8\mu} r^2
 +  {\hat V}_{h-e}(r)
                 -\frac{\left( e \mathbf{B} \times  \mathbf{K} \right)\cdot \mathbf{r}}{M} ,\nonumber
\end{equation}
where $\mu=m_e^* m_h^*/(m_e^* +m_h^*)$, $M= m_e^*+ m_h^*$, and $\rho =  m_e^*/m_h^*$  are the exciton reduced mass, total mass, and ratio of masses, respectively;  $m_e^*$ and $m_h^*$ are the effective masses of electron and hole; $e$ is the elementary charge with the positive value.  The last term in the above Hamiltonian is the motional Stark potential with the pseudomomentum $\mathbf{K}$ of the c.m. related to the temperature of exciton gas \cite{comment2020, Nhat2022}. This term can be neglected for experiments in low temperature as considered in the present study. Therefore, the Schr\"{o}dinger equation for the relative motion can be written in atomic units as
\begin{eqnarray}\label{eq8}
\left\{ -\frac{1}{2} \left( \frac{\partial^2}{\partial x^2}+\frac{\partial^2}{\partial y^2}\right)  
    +\frac{1}{8}\gamma^2 (x^2+y^2) + {\hat V}_{h-e}(r)\right.\qquad\nonumber\\
\left. \qquad+ \frac{1-\rho}{1+\rho}\,\frac{m}{2} \gamma  -E\right\} \psi(x, y) = 0,
\end{eqnarray}
where $r=\sqrt{x^2+y^2}$; energy $E$ and coordinates $x, y$ are given in the effective Hartree $E_h^{*}=\mu e^4/16\pi^2\varepsilon_0^2\hbar^2$ and effective Bohr radius $a_0^{*}=4\pi\varepsilon_0\hbar^2/\mu e^2$, respectively; $\gamma$ is dimensionless magnetic intensity related to the magnetic field by the equation $B=\gamma \times \mu E_h^{*}/ \hbar e$; $\hbar$ is the reduced Planck constant; $\varepsilon_0$ is the vacuum permittivity. In equation~\eqref{eq8}, the operator ${\hat l}_z$ is replaced by its eigenvalue (the magnetic quantum number $m$) because of the conservation of the angular momentum on the z axis.

The electron and hole interaction is described by the Rytova-Keldysh potential, initially established for excitons in thin films \cite{Rytova1967, keldysh1979} but applicable recently for excitons in monolayer TMDCs  such as $\textrm{MoS}_2$, $\textrm{MoSe}_2$, $\textrm{WS}_2$, $\textrm{WSe}_2$ \cite{chernikov2014, PhysE,Plechinger-2016,Stier2016-nat,Stier2016-nano,Zipfel2018,PRL2018,NAT2019,Liu2019,Chen2019-nano,Taghizadeh2019,Henriques2021}. In most studies, this potential is expressed via the Struve and Bessel functions and is thus suitable for numerical calculations only. For analytical calculations of the matrix elements in our approach, which significantly saves computational resources, we rewrite the Rytova-Keldysh potential by the Laplace transformation as
\begin{equation}\label{eq10}
{\hat V}_{h-e} (r)=- \frac{1}{ \kappa} 
\int\limits_{0}^{+\infty} \frac{dq}{ \sqrt{1+\alpha^2 q^2} }\;
 \textrm{e}^{-qr},
\end{equation}
where the dimensionless parameter $\alpha=r_0/\kappa a_0^{*}$ is used instead of the screening length $r_0$. Here, $\kappa$ is the average dielectric constant of the surrounding medium; $r_0$ is related to the $2D$ static polarizability for monolayer materials by the formula $r_0=2\pi \chi_{2D}$. 

\textit{Numerical method of solving the Schr{\"o}dinger equation --}
The Schr{\"o}dinger equation~\eqref{eq8} can be solved numerically by several methods. In the present work, we develop a numerical method based on the matrix eigenvalue equation solver of the Linear Algebra PACKage (LAPACK) \cite{Lapack} and the Feranchuk-Komarov operator method \cite{Feranchuk1982,Hoangbook2015}, where all matrix elements are calculated algebraically via the formalism of annihilation and creation operators with using the Levi-Civita transformation for two-dimensional atomic systems \cite{giang1993}. 

For this purpose, we rewrite the  Schr{\"o}dinger equation \eqref{eq8} in the algebraic form as
\begin{equation}
\label{eq23vn}
\left( -\frac{1}{8} \hat{{T}}+\frac{1}{8}\gamma^2{\hat R}^3+ \hat V - {\widetilde E}\,\hat{R}\right) |\psi \rangle = 0,
\end{equation}
where all operators have the form of annihilation and creation operators as presented in Appendix~\ref{appA}, Eqs.~\eqref{eq20} and \eqref{eq20l}. Here, we use the notation $\widetilde E= E-\frac{1-\rho}{1+\rho}\,\frac{m}{2} \gamma$. We also establish a basis set of wave vectors ${| k, m\rangle}$, Eq.~\eqref{eq25}, labeled by a free parameter $\omega$ and calculate all matrix elements with respect to the built basis set:
$\mathcal{R}_{jk}=\omega \,\langle j,m|\, {\hat R} \, {|k,m\rangle}$,
$\mathcal{T}_{jk}=\frac{1}{\omega} \,\langle j,m|\, {\hat T} \, {|k,m\rangle}$,
$ ({\mathcal R}^3)_{jk} =\omega^3 {\langle j, m|} {\hat R}^3{| k, m \rangle}$, and
$ {\mathcal V}_{jk} = {\langle j, m|}\,\omega {\hat V}{| k, m \rangle}$. Analytical expressions for these matrix elements are given in Eqs.~\eqref{eq28n}, \eqref{eq28m}, \eqref{eq29}, and \eqref{A2}.

We will find the wave vector of equation \eqref{eq23vn} in the expansion via the basis set as
\begin{equation}
\label{eq30n}
{| \psi^{(s)} \rangle}=  \sum_{k=|m|}^{s+|m|} C_{k}^{(s)} {| k, m\rangle},
\end{equation}
with $s+1$ unknown coefficients $C_k^{(s)}\; k=|m|, 1+|m|,..., s+|m|)$ needed to define. For the considered system, the angular momentum $l_z$ is conserved, so $m$ is the magnetic quantum number and fixed; only one running index $k$ remains.
In wave vector \eqref{eq30n}, we use only $s+1$ basis set vectors, so that the number $s$ can be considered an approximation order of the solutions. In practice, we will increment the $s$-order until getting the needed precision.

Plugging wave vector (\ref{eq30n}) into equation (\ref{eq23vn}) and acting to the left with $\langle j, m|,\,\,(j=|m|, 1+|m|,2+|m|, ..., s+|m|)$, we lead this equation to $s+1$ linear equations for the coefficients $C_{k}^{(s)}$ and corresponding energy ${E}^{(s)}$ as
\begin{eqnarray}\label{eq31n}
 \sum_{k=|m|}^{s+|m|} \left( -\frac{\omega^2}{8} \mathcal{T}_{jk} +\frac{\gamma^2}{8\omega^2} (\mathcal{R}^3)_{jk}
+ {\mathcal V}_{jk} \right.\qquad\qquad \nonumber\\
\left. -  {\widetilde{E}}^{(s)}\; {\mathcal{R}}_{jk} \right) C_{k}^{(s)} = 0,
\end{eqnarray}
where all matrix elements have explicit analytical expressions provided in Appendix~\ref{appA}.

Linear  equations \eqref{eq31n} can be rewritten in the $(s+1) \times (s+1) -$ matrix eigenvalue equation, where the eigenvalue is ${\widetilde{E}}^{(s)}$, while the eigenvector contains $s+1$ elements $C^{(s)}_k$. This matrix eigenvalue equation can be solved using the subroutine dsygvx.f of the LAPACK. 

\textit{Exact numerical solutions --} 
We note that equations \eqref{eq31n} are not solved for a sole quantum state but for a broad range of $s+1$ quantum states with the principle quantum number $n$ from $1$ to $s+1$, where the magnetic quantum number $m$ is fixed. Besides energies  $E_{nm}^{(s)}$, our Fortran codes also give wave functions $|\psi_{nm}^{(s)}\rangle$ calculated by the formula \eqref{eq30n} with the coefficients $C_{k}^{(s)}$.  The wave functions are normalized by the condition $\sum\limits_{j=|m|}^{s+|m|} C_{j}^{(s)} C_{j}^{(s)} = 1$.

Generally, if $\lim\limits_{s\rightarrow +\infty} { {E}^{(s)}}\rightarrow { E}$, the solving process converges and gives exact numerical solutions. However, in practice, we use a limited number of basis set functions to get the required precision. The more basis set functions are included in expansion \eqref{eq30n}, the better accuracy of the solution is obtained. However, another way to increase accuracy is by choosing the appropriate value of the free parameter $\omega$. Work \cite {PhysE} shows that convergence strongly depends on the free parameter, and there is an optimum region of this parameter where the convergence rate is highest. We confirm the same results even for the case $m\neq 0$ and implement the optimum values of $\omega$ in the Fortran codes. We have tested the codes with energies converged to 15 decimal places so that the solutions used in this work (required only three decimal digits) are considered numerically exact. Therefore, the precision of calculated exciton energies is determined only by the accuracy of the material parameters. 

Tables \ref{tab1} and \ref{tab2} present exciton energies in monolayers $\text{WSe}_2$ and $\text{WS}_2$ encapsulated by hBN slabs for the states with the principal quantum number $n\leq 5$. We provide only the $s$-state energies because recent experiments detect only $s$-state peaks in the absorption spectra. Energies for other states with $m\neq 0$ are available upon request. In our calculation, the exciton reduced mass $\mu=0.190 \,m_e$, screening length $r_0=4.21$ nm, and dielectric constant $\kappa=4.34$ are taken from Table~\ref{tab4n}, retrieved by our method in Sec. \ref{Sec2}; $m_e$ is the electron mass. 

\begin{table}[H]
\caption{\footnotesize \label{tab1} Magnetoexciton energies (meV) in monolayer $\text{WSe}_2$ encapsulated by hBN slabs with $r_0$ = 4.21 nm, $\mu = 0.190\,\text{m}_e$, $\kappa=4.34$. For binding energies, add the bandgap $E_g=1.892$ eV. }
\scriptsize
\vspace{0.2cm}
\begin{ruledtabular}
\begin{tabular}{c r r r r r}
Magnetic field 	 &   \multicolumn{5}{c}{Energy (meV)}           \\
    (Tesla)           &	1s &     2s    &	3s   & 4s  & 5s   \\
\hline
0	&	-168.603	&	-38.568	&	-16.558	&	-9.133	&	-5.774	\\
2	&	-168.602	&	-38.545	&	-16.439	&	-8.757	&	-4.899	\\
4	&	-168.598	&	-38.477	&	-16.090	&	-7.732	&	-2.816	\\
6	&	-168.592	&	-38.364	&	-15.531	&	-6.238	&	-0.086	\\
8	&	-168.583	&	-38.207	&	-14.789	&	-4.409	&	3.038	\\
10	&	-168.571	&	-38.008	&	-13.887	&	-2.332	&	6.432	\\
15	&	-168.530	&	-37.329	&	-11.071	&	3.617	&	15.698	\\
20	&	-168.473	&	-36.412	&	-7.641	&	10.298	&	25.691	\\
25	&	-168.400	&	-35.282	&	-3.756	&	17.472	&	36.163	\\
30	&	-168.311	&	-33.959	&	0.476	&	25.007	&	46.979	\\
35	&	-168.206	&	-32.466	&	4.987	&	32.819	&	58.060	\\
40	&	-168.085	&	-30.818	&	9.724	&	40.854	&	69.350	\\
45	&	-167.948	&	-29.033	&	14.652	&	49.072	&	80.813	\\
50	&	-167.795	&	-27.122	&	19.741	&	57.445	&	92.420	\\
60	&	-167.445	&	-22.970	&	30.321	&	74.570	&	115.987	\\
70	&	-167.034	&	-18.438	&	41.336	&	92.101	&	139.931	\\
80	&	-166.565	&	-13.584	&	52.700	&	109.954	&	164.172	\\
90	&	-166.039	&	-8.454	&	64.349	&	128.069	&	188.654	\\

\end{tabular}
\end{ruledtabular}
\end{table}

\begin{table}[H]
\caption{\footnotesize \label{tab2} Magnetoexciton energies (meV) in monolayer $\text{WS}_2$ encapsulated by hBN slabs with $r_0$ = 3.76 nm, $\mu = 0.175\,\text{m}_e$, $\kappa=4.16$. For binding energies, add the bandgap $E_g=2.238$ eV. }
\scriptsize
\vspace{0.2cm}
\begin{ruledtabular}
\begin{tabular}{c r r r r r}
Magnetic field 	 &	    \multicolumn{5}{c}{Energy (meV)}           \\
    (Tesla)           &	1s &     2s    &	3s   & 4s  & 5s   \\
\hline
0	&	-178.617	&	-39.725	&	-16.899	&	-9.282	&	-5.853	\\
2	&	-178.616	&	-39.699	&	-16.763	&	-8.851	&	-4.856	\\
4	&	-178.611	&	-39.623	&	-16.367	&	-7.690	&	-2.517	\\
6	&	-178.604	&	-39.496	&	-15.735	&	-6.012	&	0.517	\\
8	&	-178.594	&	-39.320	&	-14.899	&	-3.972	&	3.972	\\
10	&	-178.581	&	-39.096	&	-13.887	&	-1.665	&	7.714	\\
15	&	-178.537	&	-38.336	&	-10.743	&	4.910	&	17.893	\\
20	&	-178.475	&	-37.313	&	-6.935	&	12.264	&	28.842	\\
25	&	-178.395	&	-36.054	&	-2.640	&	20.139	&	40.295	\\
30	&	-178.297	&	-34.584	&	2.028	&	28.396	&	52.113	\\
35	&	-178.183	&	-32.928	&	6.992	&	36.948	&	64.210	\\
40	&	-178.050	&	-31.104	&	12.198	&	45.735	&	76.529	\\
45	&	-177.901	&	-29.130	&	17.605	&	54.716	&	89.030	\\
50	&	-177.735	&	-27.020	&	23.185	&	63.860	&	101.683	\\
60	&	-177.352	&	-22.445	&	34.770	&	82.551	&	127.364	\\
70	&	-176.904	&	-17.460	&	46.818	&	101.671	&	153.442	\\
80	&	-176.392	&	-12.128	&	59.234	&	121.131	&	179.834	\\
90	&	-175.817	&	-6.500	&	71.955	&	140.869	&	206.482	\\

\end{tabular}
\end{ruledtabular}
\end{table}

We consider the magnetic field intensity up to 90 Tesla only because of the current laboratory limit in generating the magnetic field. Indeed, most studies deal with the intensity from 30 to 65 Tesla \cite{Plechinger-2016, Stier2016-nat, Stier2016-nano, Zipfel2018, PRL2018, Chen2019-nano, Liu2019}, while the highest intensity recently achieved is 91 Tesla \cite{NAT2019}. 
Also, for binding energies, we need to subtract the bandgap (extracted from experimental exciton energies in Table~\ref{tab4n}) from the calculated exciton energies.  

\section{\label{Sec2}Retrieval of material properties from energy spectra}

\begin{figure}[htbp]
\begin{center}
\includegraphics[width=0.75 \columnwidth]{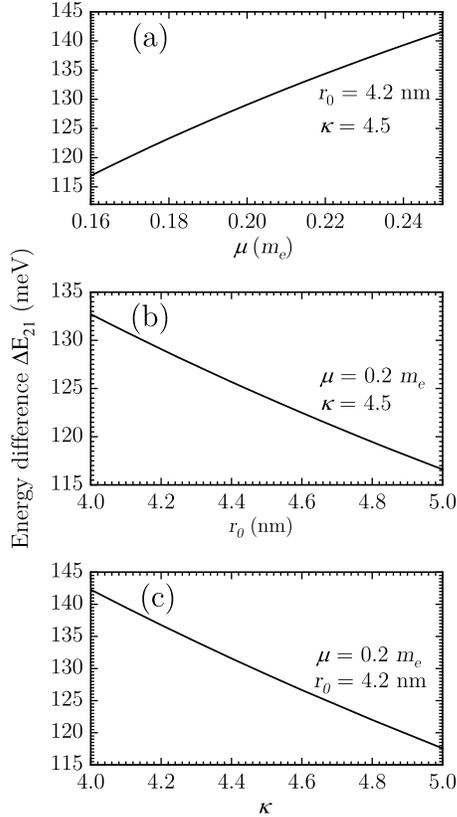}
\caption{Sensitivity of the exciton energy difference $\Delta E_{21} = E_{2s}-E_{1s}$ on the exciton reduced mass (a), screening length (b), and average dielectric constant of the surrounding medium (c). }
\label{fig2}
\end{center}
\end{figure}

\textit{Sensitivity of exciton energies on material parameters --} There are four parameters in the Schr{\"o}dinger equation \eqref{eq8}  of an exciton in a monolayer TMDC that vary for different materials. They are the exciton reduced mass $\mu$, screening length $r_0$ (related to the $2D$ polarizability), average dielectric constant $\kappa$ of the surrounding medium, and the mass ratio $\rho$. We consider only the 
$s$-states, so the mass ratio $\rho$ disappears in the equation. Remain only three material parameters ($\mu$, $r_0$, and $\kappa$) needed to retrieve. Therefore, we now investigate the sensitivity of exciton energies on these parameters and show the results in Fig.~\ref{fig2}.

From our calculations, Figs.~\ref{fig2} (a), (b), and (c) present the energy difference $\Delta E_{21}=E_{2s}-E_{1s}$ dependent on $\mu$, $r_0$, and $\kappa$, respectively, for monolayer TMDCs. The changes are 24.4 meV (18\%), -16.1 meV (-12\%), and -25.1 meV (-19\%), respectively, when varying exciton reduced mass from $0.16$ to $0.25\, m_e$, screening length from 4.0 to 5.0 nm, and dielectric constant from 4.0 to 5.0. Analogically for the energy difference  $\Delta E_{32}=E_{3s}-E_{2s}$ (not shown in the figure), the changes are 6.5 meV (30\%), - 1.4 meV (- 6\%), and - 6.8 meV (- 31\%), respectively. On the other hand, the measurement accuracy for exciton energies in the hBN environment is less than 1.0 meV, so the energy changes are significant enough for the experimental detection. Therefore, we conclude that exciton energies are sensitive to the change of reduced mass, screening length, and dielectric constant and will use this fact for developing our extraction method. 

\textit{Fitting method for exciton reduced mass, screening length, and dielectric constant --} 
The work of Stier \textit{et al.} (2018) \cite{PRL2018} for exciton energies in monolayer $\text{WSe}_2$ encapsulated by hBN slabs with $\kappa=4.5\,$ provides experimental data of 130.0 meV for energy difference $\Delta E_{21}$ and 22.0 meV for $\Delta E_{32}$. This work also performs the theoretical calculation with 124.0 meV and 21.3 meV respectively for the mentioned energy differences. The discrepancies between experimental data and theoretical calculation are 4.0 \% and 3.2 \%, which we attribute to the inaccuracy of the material parameters $\mu$, $r_0$, and $\kappa$ used in the calculation. The sensitivity of exciton energies on the material parameters inspires us to find the values of the reduced mass $\mu$, screening length $r_0$, and dielectric constant $\kappa$ so that the theoretical results best fit the experimental data.

\begin{figure}[H]
\begin{center}
\includegraphics[width=0.95 \columnwidth]{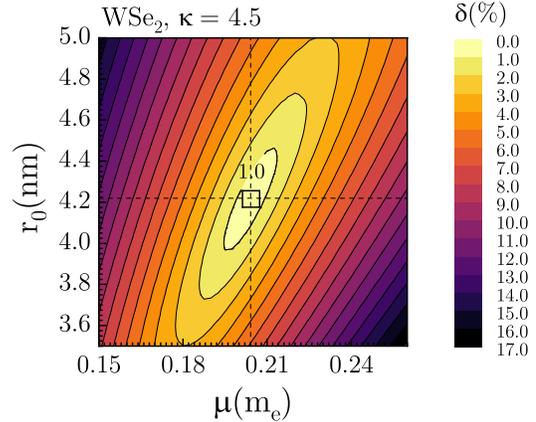}
\caption{Relative discrepancy between the experimental data for monolayer $\text{WSe}_2$ \cite{PRL2018} and theoretical energy differences $\Delta E_{21}$ and $\Delta E_{32}$, calculated with varied exciton reduced mass $\mu$, screening length $r_0$, and fixed $\kappa = 4.5$. There is a minimum at $\mu=0.204\, \text{m}_e$ and $r_0=4.21$ nm. }
\label{fig3}
\end{center}
\end{figure}

Figure~\ref{fig3} shows the relative discrepancy between the experimental data from Ref.~\cite{PRL2018} and the theoretical energy differences. We calculate it by the formula 
\begin{eqnarray}\label{eq2-1}
\delta=\frac{1}{2}\left( \frac{|\Delta E^{\text {theo}}_{21}-\Delta E^{\text{exp}}_{21}|}{\Delta E^{\text{exp}}_{21}}
      +\frac{|\Delta E^{\text {theo}}_{32}-\Delta E^{\text{exp}}_{32}|}{\Delta E^{\text{exp}}_{32}}\right) 
\end{eqnarray}
varying the exciton reduced mass $\mu$ and screening length $r_0$ by the steps $\Delta \mu = 0.0025 \, m_e$ and $\Delta\, r_0 = 0.025$ nm while fixing the value $\kappa=4.5$. There is a minimum discrepancy at $\mu=0.204 \, m_e$ and $r_0=4.21$ nm, which gives true values for the exciton reduced mass and screening length ($2D$ polarizability) of the considered monolayer $\text{WSe}_2$.

\begin{figure}[H]
\begin{center}
\includegraphics[width=0.95\textwidth]{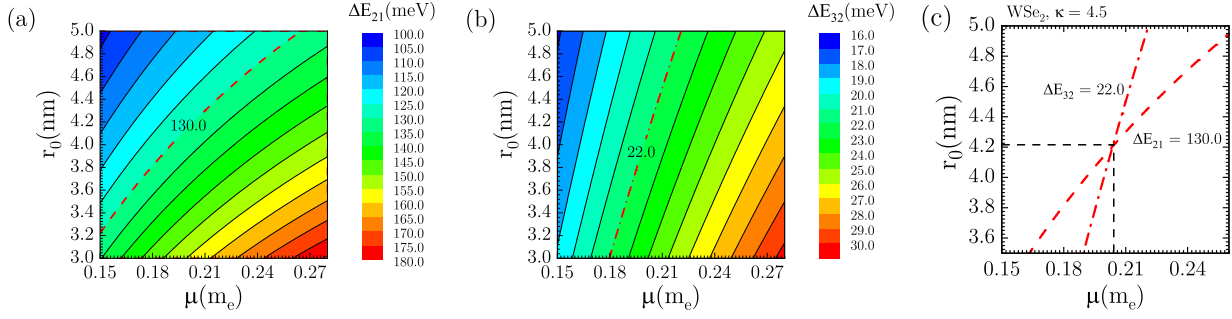}
\caption{Sensitiveness of the exciton energy differences (a) $\Delta E_{21} = E_{2s}-E_{1s}$ and (b) $\Delta E_{32}=E_{3s}-E_{2s}$ on the exciton reduced mass $\mu$ and the screening length $r_0$. The energies are calculated for $\kappa = 4.5\,$. (c) The intersection of two lines ($\Delta E_{21}=130.0$ meV and $\Delta E_{32}=22.0$ meV) gives the finding values of  $\mu$ and $r_0$. }
\label{fig4}
\end{center}
\end{figure}

Mathematically, the minimum in Fig.~\ref{fig3} can be understood because there are two constraints ($\Delta E_{21}$ and $\Delta E_{32}$) for two parameters ($\mu$ and $r_0$) to be defined. However, we also provide a more comprehensible explanation demonstrated in Fig.~\ref{fig4}. Panel (a) presents the energy difference $\Delta E_{21}$ dependent on $\mu$ and $r_0$, which is not mono-semantic. Each energy difference value corresponds to a set of values $\mu$ and $r_0$, establishing a curved line in the diagram. Analogically, Panel (b) shows a similar picture -- each energy difference value $\Delta E_{32}$ corresponds to a curved line in the plane ($\mu$, $r_0$). As shown in Panel (c), the two lines ($\Delta E_{21}=$130.0 meV and $\Delta E_{32}=$ 22.0 meV) intersect at one point,  defining the material parameters for monolayer $\text{WSe}_2$, $\mu=0.204 \, \text{m}_e$ and $r_0= 4.21$ nm, consistent with the results shown in Fig.~\ref{fig3}.

Work~\cite{PRL2018} also provides exciton energy spectra dependent on the magnetic intensity. We can use this information to get a more precise value of the dielectric constant $\kappa$ of the surrounding medium (hBN in this case). First, we change $\kappa$ around the value 4.5, from 4.0 to 5.0, and for each value, we get the optimum values of $\mu$ and $r_0$ by the above procedure. The results presented in Table~\ref{tab3} show that the screening length $r_0$ does not change but is around the value of $4.21$. For each pair of optimum values of $\mu$ and $\kappa$, we calculate energies for $1s$, $2s$, $3s$, and $4s$ states of the exciton at the magnetic intensity for which the experimental energies are available in Ref.~\cite{PRL2018}. By the least square method, we get the values of $\mu=0.190\, \text{m}_e$ and $\kappa=4.34$, where the theoretical energies best fit the experimental data. Here, we note that the screening effect in monolayer TMDC is the consequence of dimensionality reduction, that is why the screening length $r_0$ in Table \ref{tab3} is almost independent of the dielectric constant $\kappa$. In contrast, the exciton reduced mass $\mu$ strongly depends on $\kappa$. However, this fact needs more careful investigation with analytical exciton energies as functions of material parameters. Some works are available for analytical energies of exciton in monolayer TMDCs \cite{Molas2019, Hieu2022}; however, they are unsuitable for our analysis. Therefore, we left it for further study.

\onecolumngrid

\twocolumngrid

\begin{table}[H]
\caption{\footnotesize \label{tab3} Optimum values of exciton reduced mass $\mu$ and screening length $r_0$ extracted with different values of dielectric constant $\kappa$ for monolayer WSe$_2$ encapsulated by hBN slabs. }
\scriptsize
\vspace{0.2cm}
\begin{ruledtabular}
\begin{tabular}{l l l }

Dielectric constant   &	 Exciton reduced mass  &	  Screening length $r_0$ 	\\
	   \quad   $\kappa$               &		   	\quad$ \mu$ ($m_e$)	            &	\quad $r_0$ (nm)	     \\
\hline
\quad 5.0 & 	\quad 0.252 &	\quad  4.208		  	\\
\quad 4.8 &	 \quad 0.232	&	\quad 4.208 \\
	\quad 4.6 &	 \quad 0.213	&	\quad 4.208 \\
\quad 4.5 &	 \quad 0.204	&	\quad 4.208 \\
\quad 4.4 &	 \quad 0.195	&	\quad 4.209 \\
\quad 4.35 &	 \quad 0.191	&	\quad 4.209 \\
\quad 4.34 & \quad 0.190  & \quad 4.209 \\
\quad 4.33 & \quad 0.189  & \quad 4.209 \\
\quad 4.3 &	 \quad 0.186	&	\quad 4.209 \\
\quad 4.2 &	 \quad 0.178	&	\quad 4.209 \\
\quad 4.1 &	 \quad 0.169	&	\quad 4.207 \\
\quad 4.0 &	 \quad 0.161	&	\quad 4.209 \\

\end{tabular}
\end{ruledtabular}
\end{table}

\begin{figure}[htbp]
\begin{center}
\includegraphics[width=0.90\columnwidth]{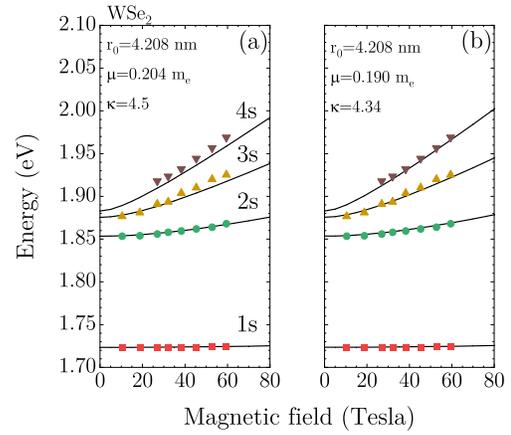}
\caption{Magnetoexciton energy spectra calculated with different values of material parameters for monolayer WSe$_2$: (a) $\mu=0.204\, \text{m}_e$, $r_0=4.208$ nm, and $\kappa=4.5$; (b) $\mu=0.190\, \text{m}_e$, $r_0=4.208$ nm, and $\kappa=4.34$. The results in (b) agree better with the experimental data of Ref.~\cite{PRL2018}, indicated by the color symbols. For the binding energies in the figures, the bandgap $E_g=1.892$ eV is used. }
\label{fig5}
\end{center}
\end{figure}

For illustration, we present in Fig.~\ref{fig5} exciton energy spectra calculated for two sets of $\mu$, $r_0$, and $\kappa$, compared with the experimental data (color symbols). It is clear that the theoretical spectrum best fits the experimental data at the optimum values $\mu$, $r_0$, and $\kappa$. We note that the bandgap energy $E_g= 1.892$ eV for monolayer WSe$_2$ in Fig.~\ref{fig5} is chosen so that the calculated binding $1s$ exciton energy equals the experimental one. We also extracted the bandgap for monolayer WS$_2$, given in Table~\ref{tab4n}.

\textit{Extracted fundamental optoelectronic material parameters for monolayer TMDCs --} 
The method suggested above can retrieve the reduced mass, screening length, and dielectric constant of any monolayer TMDC from the measured energy differences $\Delta E_{21}$ and $\Delta E_{32}$ combined with the magnetoexciton energy spectra. For this task, we have Fortran codes available upon request. Here, we demonstrate the method for the experimental data extracted from Ref.~\cite{PRL2018} for monolayer $\text{WSe}_2$:  $\Delta E_{21}$= 130.0 meV, $\Delta E_{32}$= 22.0 meV; and  from Ref.~\cite{NAT2019} for monolayer $\text{WS}_2$: $\Delta E_{21}$= 139.2 meV, $\Delta E_{32}$= 22.1 meV. The retrieved exciton reduced mass $\mu$, screening length $r_0$, and dielectric constants $\kappa$ are given in Table~\ref{tab4n} compared with data from other works. The free-particle bandgaps are also obtained by fitting the experimental and calculated $1s$ energies. For reference, we also calculate the diamagnetic coefficient $\sigma$ and exciton radii $r_{1s}$, $r_{2s}$, and $r_{3s}$ for $1s$, $2s$, and $3s$ states presented in the Table. 

We now discuss our results in comparison with other works. First, for the free-particle bandgap, we retrieve it by correlating the theoretical and experimental energies as it was performed in Refs.~\cite{PRL2018,NAT2019}. However, the theoretical ones are numerically exact in our calculation while approximated by the variational method in these cited references, resulting in the difference between the bandgaps of 1.892 eV (Present work) and 1.890 eV (Ref.~\cite{PRL2018}) for monolayer WSe$_2$. Meanwhile, our result and Ref.~\cite{NAT2019} are the same, $E_g=2.238$ eV. 
We note that these bandgaps revealed from exciton absorption peaks are smaller than those obtained by the GW calculation (WSe$_2$: 2.100 eV and WS$_2$: 2.530 eV) and bigger than those calculated by the DFT method (WSe$_2$: 1.730 eV and WS$_2$: 2.050 eV), based on the computational 2D materials database \cite{haastrup2018, gjerding2021}. Compared with the direct measurements by the scanning tunneling spectroscopy, the extracted bandgap for monolayer WS$_2$ agrees well with the experiments (2.238 vs. 2.140 eV) \cite{jo2014}, but the one for WSe$_2$ is underestimated (1.892 vs. 2.080 eV) \cite{zhang2015,chiu2015}.

\onecolumngrid

\begin{table}[H]
\caption{\footnotesize \label{tab4n} Fundamental optoelectronic material parameters (exciton reduced mass $\mu$, screening length $r_0$, dielectric constant $\kappa$, and bandgap energy $E_g$) extracted in the present work compared with data in other references. Also some exciton properties (diamagnetic coefficient $\sigma$ for the $1s$ state, exciton radii $r_{1s}$, $r_{2s}$, and $r_{3s}$ calculated with the extracted material parameters. }
\scriptsize
\vspace{0.2cm}
\begin{ruledtabular}
\begin{tabular}{l l l l l l l l l l}

Material	&	  $\mu$ &	  $r_0$ &	$\kappa$   &	$E_g$&$\sigma$ & $r_{1s} $ 	& $r_{2s} $ &$r_{3s} $ &References	\\
	          &	($m_e$)  &	 (nm)&    &	(eV)& ($\mu$eV/T$^2$)&(nm) 	&(nm)	 &(nm)  & 	\\
\hline
&		   &		  &		  &		  &		  &		&   &		&\\
WSe$_2$	 &	0.190	&	4.21 &4.34	&1.892	&0.28 &1.68  & 	7.01 	&16.09& Present work	\\
		&	    0.20	   &	4.5	&4.5&	1.890&0.31 &1.7	&	 6.6	& 14.3 &\cite{PRL2018}\\
        & 0.20 	&	5.0&3.97	  &1.884&	0.24	&1.6	&	  8.24 &	17.0 &   \cite{Liu2019}    \\
	&	0.22 	&	4.51 &4.5& 1.900	&	0.25 & 1.6	&	6.5	 &	14.7	&\cite{Chen2019-nano}\\
	&	0.22 &	 4.5 &3.3 &---&	0.32 &1.79	&	--- & --- &\cite{Stier2016-nano}	\\
\hline
	&		   &		  &		  &		  &		  &			&	\\

WS$_2$	&	0.175 & 3.76&4.16 & 2.238&0.34&1.69	&	7.13 & 16.49&Present work	\\
	&	0.175 &	3.4 &4.35&2.238& 0.4  &1.8	&	---	&--- &\cite{NAT2019}\\
	&	0.15 	&	4.0& 4.5 &---&---	&2.45	&	--- &--- &\cite{Zipfel2018}	\\
	&	0.16 &	5.3&1.0	&---&	0.32 &1.53	&	---	&--- &\cite{Stier2016-nat}\\
	&	0.15 &	---&1.55	&---&	0.90 &2.5	&	--- & ---&\cite{Plechinger-2016}	\\
	&		&		      &		&		  &		  &	&&	&		\\
\end{tabular}
\end{ruledtabular}
\end{table}

\twocolumngrid

Concerning the exciton reduced mass $\mu$ for WSe$_2$, our extracted result is close to the one of Ref~\cite{PRL2018}, i.e., 0.19 versus 0.20 $m_e$. The discrepancy is because of the difference in the fitting schemes used in the two works. Work~\cite{PRL2018} extracts only one parameter $\mu$ while estimating the screening length $r_0=4.5$ nm from previous theoretical studies and experimental measurements \cite{berkelbach2013}. Meanwhile, the dielectric constant $\kappa$ in Ref.~\cite{PRL2018} is also taken from the infrared measurements \cite{geick1966}. In contrast, we consider $\mu$, $r_0$, $\kappa$, and $E_g$ as material parameters and extract them from the magnetoexciton energies. From Table~\ref{tab4n}, our extracted parameters $r_0=4.21$ nm and $\kappa=4.34$ are close to the previous experimental and theoretical data. Particularly, calculated from the extracted screening length $r_0=4.21$ nm by the equation $\chi_{2D}=r_0/2\pi$, the $2D$ polarizability is 6.7 \AA, very close to the DFT calculation (7.18 \AA \cite{berkelbach2013} and 6.72 \AA \cite{haastrup2018, gjerding2021}).  

We note that using the high-$B$ shifts of the $3s/4s$ excitons in Ref.~\cite{PRL2018} to retrieve the exciton reduced mass leads to a wide range for it, from 0.16 to $0.23 \,m_e$. This inaccuracy is because the electron-hole interaction is negligible only when the dimensionless magnetic intensity $\gamma$ is larger than 10 \cite{Stafford1990, KOCH1999}. This condition is equivalent $a_0^*> 3\,l_B$, noticing that $\gamma = (a_0^*/l_B)^2$ where $l_B = \sqrt{\hbar/eB}$ is a magnetic length. Meanwhile, for WSe$_2$ with $\mu \sim 0.20 \,m_e$ at $B = 65$ Tesla, it has $a_0^*\sim 0.08 \,l_B$ only. Nevertheless, Work~\cite{PRL2018} has also modeled the exciton by the Rytova-Keldysh potential and calculated the diamagnetic shifts in the magnetic field range up to 65 Tesla, and with the numerically acceptable magnetoexciton energies, it has retrieved a good result for $\mu$ (about $0.20 \, m_e$) despite the current laboratory limit in the magnetic field generation.

For monolayer WS$_2$, we compare our results with those of Ref.~\cite{NAT2019}. For the exciton reduced mass $\mu$, this reference uses the $4s/5s$ exciton states where the material parameters such as $r_0$ and $\kappa$ are supposed to influence the energies weakly at the strong magnetic field. The obtained exciton reduced mass of $\mu=0.175\, m_e$ exactly coincides with ours. Then, Work~\cite{NAT2019} modeled excitons by the Rytova-Keldysh potential with different values of screening length $r_0$ and dielectric constant $\kappa$ at the weak magnetic intensity to fit the calculated binding energies with the experimental data. As a result, they get $r_0=3.4$ nm and $\kappa=4.35$. Our fitting scheme is different and with numerically exact exciton energies used. As shown in Table~\ref{tab4n}, our result for the screening length, $r_0=3.76$ nm, is bigger than the one of Ref.~\cite{NAT2019}, about $10\%$, and closer to the DFT calculation. Indeed, we calculated the $2D$ polarizability related to the screening length with the result $\chi_{2D}=5.98 \text{ \AA}$ which is close to the DFT estimation (6.03 \AA  \cite{berkelbach2013}, 6.393 \AA \cite{kylanpaa2015}, and 5.9 \AA  \cite{haastrup2018, gjerding2021}).

\section{\label{conc}Conclusion}

We have shown the sensitivity of the exciton energy differences among three exciton quantum states ($1s$, $2s$, and $3s$) in the monolayer TMDCs to the material properties. It inspires us to propose a method to retrieve the exciton reduced mass, screening length (related to the $2D$ polarizability), and dielectric constant of the surrounding medium from experimental magnetoexciton energies available recently. Applying the proposed method to monolayers $\text{WSe}_2$ and $\text{WS}_2$, we have obtained results  for the material properties that complement well the available data. The method could be extended for other monolayer TMDCs, such as $\text{MoS}_2$, $\text{MoSe}_2$, and $\text{MoTe}_2$, which are the subject of recent intensive investigation. Also, the mass ratio $\rho=m_e^*/m_h^*$ is an important material property that needs to be extracted. Our approach can be applied for this purpose required experimental energies of states with $m \neq 0$, which could be obtained from nonlinear optical response or thermoinduced magnetoexciton peaks in linear optical response.

For the mentioned-above investigation, we have developed an effective method for solving the Schr{\"o}dinger equation of a magnetoexciton in a monolayer TMDC. The method gives a very fast and convergent procedure to get highly accurate magnetoexciton energies and wave functions suitable for the fitting method, which usually requires a huge data generation. Besides, all matrix elements for the Hamiltonian are obtained in analytical expressions that may be useful for further investigation of analytical magnetoexciton energies as functions of material parameters. Fortran codes for magnetoexciton energy spectra in monolayer TMDCs are available upon request and will be published elsewhere.

\begin{acknowledgments}
D.-N.Ly and N.-H.P. are funded by Ho Chi Minh City University of Education Foundation for Science and Technology under grant numbers CS.2019.19.43TD and CS.2019.19.44TD. This work is funded by Foundation for Science and Technology of Vietnam Ministry of Education and Training under grant number B2022-SPS-09-VL. This work was carried out by the high-performance cluster at Ho Chi Minh City University of Education, Vietnam.
\end{acknowledgments}

\appendix
\section{Analytical matrix elements} \label{appA}

For more effectively solving the Schr{\"o}dinger equation \eqref{eq8}, we first rewrite it in the $(u,v)$ space by the Levi-Civita transformation $x=u^2-v^2,\; y=2uv$, where the interaction potential in the $(u, v)$ space is defined as ${\hat V}(u, v)=\left( {u}^2+{v}^2\right) {\hat V}_{h-e}$. The distance and angular momentum have the compact form $r=u^2+v^2$ and
${\hat l_z}= -\frac{i}{2}\left( v\frac{\partial}{\partial u} - u\frac{\partial}{\partial v} \right)$. More about the application of the Levi-Civita transformation to two-dimensional atomic systems can also be found in Ref.\cite{giang1993}. 

One advantage of using the equation in $(u, v)$ space is to apply the algebraic formalism via annihilation and creation operators $\hat a (\omega)$, $\hat a^{+} (\omega)$, $\hat b (\omega)$, and $\hat b^{+} (\omega)$, where the calculation technique is based on the commutation relations 
	$\left[{\hat a}, {\hat a}^{+} \right]=1,\;\;\left[{\hat b}, {\hat b}^{+} \right]=1$, and the basis vectors can be presented in the form
\begin{equation}\label{eq25}
          {| k,m \rangle} = \frac{1}{\sqrt{{(k+m)}! {(k-m)}!}} 
            ({\hat a}^{+})^{k+m} ({\hat b}^{+})^{k-m}| 0 (\omega) \rangle
\end{equation}
with the vacuum state $|0 (\omega)\rangle$ defined by the equations
      $ {\hat a}\,{| 0 (\omega) \rangle} = 0,\;\; {\hat b} \,| 0 (\omega) \rangle =0.$
Here, the running quantum numbers have values $m=0,\pm 1,\pm 2, \ldots$ and  $k=|m|, 1+|m|, 2+|m|, \ldots$.

Using the annihilation and creation operators, we can rewrite all the terms in the Schr{\"o}dinger equation as
\begin{eqnarray}\label{eq20}
{\hat T}=&\dfrac{\partial^2}{\partial {u}^2} +\dfrac{\partial^2}{\partial {v}^2} &= \;\omega\left({\hat a}{\hat b} 
    + {{\hat a}}^{+}{{\hat b}}^{+} - {{\hat a}}^{+}{\hat a} - {{\hat b}}^{+}{\hat b}  - 1 \right),\nonumber\\
{\hat R}=&{u}^2+{v}^2 &= \;\frac{1}{\omega} \left( {\hat a}{\hat b} + {{\hat a}}^{+} {{\hat b}}^{+} + {{\hat a}}^{+}{\hat a} + {{\hat b}}^{+}{\hat b} + 1 \right).\nonumber\\
\end{eqnarray}
Particularly, the interaction potential can be rewriten as
\begin{eqnarray}\label{eq20l}
{\hat V}(u,v)=- \frac{1}{ \kappa} 
\int\limits_{0}^{+\infty} \frac{dq}{ \sqrt{1+\alpha^2 q^2} }
 \textrm{e}^{-q {\hat R}}{\hat R}.
\end{eqnarray}

With the algebraic forms \eqref{eq20} and \eqref{eq20l}, we can easily calculate all matrix elements just using the commutation relations of the annihilation and creation operators. Detailed calculation method can be found in monograph \cite{Hoangbook2015}. Here, in this Appendix, we provide only the results for the matrix elements. They are as follows  
\begin{eqnarray}\label{eq28n}
\mathcal{R}_{jk}&=&\omega \,\langle j,m|\, {\hat R} \, {|k,m\rangle} =
\sqrt{k^2-m^2} \,\delta_{j, k-1} \nonumber\\
&& +(2k+1)\, \delta_{jk}  +\,\sqrt{(k+1)^2 - m^2} \,\delta_{j, k+1}\,,\qquad
\end{eqnarray}
\begin{eqnarray}\label{eq28m}
\mathcal{T}_{jk}&=&\frac{1}{\omega} \,\langle j,m|\, {\hat T} \, {|k,m\rangle}= 
\sqrt{k^2-m^2} \,\delta_{j, k-1} \nonumber \\
&& -(2k+1)\, \delta_{jk}+\,\sqrt{(k+1)^2 - m^2} \,\delta_{j, k+1}\,,\qquad
\end{eqnarray}
\begin{eqnarray}\label{eq29}
&({\mathcal R}^3)_{jk} = \omega^3 \langle j,m| \,{\hat R}^3\, {|k,m\rangle} 
      \qquad\qquad\quad\quad\qquad\qquad&\nonumber\\
     &=2\,(5 k^2+5k+3-3m^2)(2k+1)\,\delta_{jk} \quad&\nonumber\\
     &+ 3\,(5k^2+1-m^2) \sqrt{k^2-m^2} \,\delta_{j, k-1}&\nonumber\\
     &+ 3\,(2k-1)\sqrt{k^2-m^2}\sqrt{(k-1)^2-m^2} \,\delta_{j, k-2}&\nonumber\\
     &+ \sqrt{k^2-m^2} \sqrt{(k-1)^2-m^2}\sqrt{(k-2)^2-m^2} \,\delta_{j, k-3}&\nonumber\\
    &+3\,(5k^2+10k+6-m^2)\sqrt{(k+1)^2-m^2} \,\delta_{j, k+1}\quad\;&\nonumber\\
    & +3\,(2k+3)\sqrt{(k+1)^2-m^2}\sqrt{(k+2)^2-m^2} \,\delta_{j, k+2}&\nonumber\\
    &+\sqrt{(k+1)^2-m^2}\sqrt{(k+2)^2-m^2}\quad&\nonumber\\
    &\times \sqrt{(k+3)^2-m^2} \,\delta_{j, k+3}\;\;&
\end{eqnarray}
Here, we use the Kronecker delta $\delta_{jk}$.
 
Differently, it is not trivial to calculate matrix elements of the operator $\hat V$. However, by using the technique of constructing operators in a normal form of annihilation and creation operators, given in Ref.~\cite{Hoangbook2015} (pages 232-233), we have formula
\begin{eqnarray}\label{A4}
 \textrm{e}^{-q {\left( {\hat a}{\hat b} + {{\hat a}}^{+} {{\hat b}}^{+} + {{\hat a}}^{+}{\hat a} + {{\hat b}}^{+}{\hat b} + 1 \right)}}
=\textrm{e}^{-\frac{q}{1+q}\,{{\hat a}}^{+} {{\hat b}}^{+}}\nonumber\\
\times \,  \textrm{e}^{-\ln(1+q)\,\left( {\hat a}^{+}{\hat a} + {\hat b}^{+}{\hat b} + 1 \right)}
                     \textrm{e}^{-\frac{q}{1+q}\,{\hat a}{\hat b}}.
\end{eqnarray}
With this operator in this normal form, we can apply the algebraic technique to get 
\begin{eqnarray}\label{A2}
     { \mathcal V}_{jk} &=& \langle j, m|\omega {\hat V}{| k, m \rangle}\nonumber\\
    & =& (2k+1)\, U_{jk} +\sqrt{k^2-m^2}\, U_{j, k-1} \nonumber\\
       &&\quad\quad+\sqrt{(k+1)^2-m^2}\, U_{j, k+1}
\end{eqnarray}
with 
\begin{eqnarray}\label{A7}
 U_{jk} &=& -\frac{1}{\kappa\,\alpha} \sum_{s=|m|}^{\text{min}(k,j)} \sum_{t=0}^{j+k-2s} (-1)^{j+k+t}
    {{j+k-2s} \choose {t}} \nonumber\\
         & \times& \sqrt{ {{j+m}\choose{s+m}}} \sqrt{{{j-m}\choose {s-m}}} 
               \sqrt{{{k+m}\choose {s+m}}} \sqrt{{{k-m}\choose {s-m}}}  \nonumber\\
&\times& \int\limits_{0}^{+\infty} \frac{dq}{(1+q)^{2s+t+1}\sqrt{q^2+1/\omega^2\alpha^2}}\,,
\end{eqnarray}
where ${n \choose k} = \frac{n!}{(n-k)!k!}$ is a binomial coefficient.

In Eq. (\ref{A7}), the definite integrals 
\[ J_p (x)=\int\limits_{0}^{+\infty} \frac{dq}{(1+q)^p\sqrt{q^2+x^2}}\]
with $p \geq 1$ and $x=1/\omega \alpha>0$ are easy to calculate numerically. Besides, for an analytical formulation, we can derive an iterative formula for these integrals as follows
\begin{equation}\label{A8}
J_p=\frac{(2p-3)J_{p-1}-(p-2)J_{p-2}+x}{(x^2+1)(p-1)}
\end{equation}
for $p \geq 2$, where $J_1(x)$ has the following explicit formula
\[J_1 (x)=\frac{\ln{\left(x+\sqrt{x^2+1}\right)}+\ln{\left(1+\sqrt{x^2+1}\right)}-\ln(x) }{\sqrt{x^2+1}}.\]
Noting that althought $J_0(x)$ is disvergent, relation \eqref{A8} is still valid for $p=2$ by considering the limit
\[ {\textrm{lim}}_{p \rightarrow 0} \;pJ_p(x)=1\]
so that
\begin{eqnarray}
J_2(x)&& =\frac{J_1(x)-1+x}{x^2+1}\,. \nonumber
\end{eqnarray}

\bibliography{apssamp}

\begin{thebibliography}{44}%
\makeatletter
\providecommand \@ifxundefined [1]{%
 \@ifx{#1\undefined}
}%
\providecommand \@ifnum [1]{%
 \ifnum #1\expandafter \@firstoftwo
 \else \expandafter \@secondoftwo
 \fi
}%
\providecommand \@ifx [1]{%
 \ifx #1\expandafter \@firstoftwo
 \else \expandafter \@secondoftwo
 \fi
}%
\providecommand \natexlab [1]{#1}%
\providecommand \enquote  [1]{``#1''}%
\providecommand \bibnamefont  [1]{#1}%
\providecommand \bibfnamefont [1]{#1}%
\providecommand \citenamefont [1]{#1}%
\providecommand \href@noop [0]{\@secondoftwo}%
\providecommand \href [0]{\begingroup \@sanitize@url \@href}%
\providecommand \@href[1]{\@@startlink{#1}\@@href}%
\providecommand \@@href[1]{\endgroup#1\@@endlink}%
\providecommand \@sanitize@url [0]{\catcode `\\12\catcode `\$12\catcode
  `\&12\catcode `\#12\catcode `\^12\catcode `\_12\catcode `\%12\relax}%
\providecommand \@@startlink[1]{}%
\providecommand \@@endlink[0]{}%
\providecommand \url  [0]{\begingroup\@sanitize@url \@url }%
\providecommand \@url [1]{\endgroup\@href {#1}{\urlprefix }}%
\providecommand \urlprefix  [0]{URL }%
\providecommand \Eprint [0]{\href }%
\providecommand \doibase [0]{https://doi.org/}%
\providecommand \selectlanguage [0]{\@gobble}%
\providecommand \bibinfo  [0]{\@secondoftwo}%
\providecommand \bibfield  [0]{\@secondoftwo}%
\providecommand \translation [1]{[#1]}%
\providecommand \BibitemOpen [0]{}%
\providecommand \bibitemStop [0]{}%
\providecommand \bibitemNoStop [0]{.\EOS\space}%
\providecommand \EOS [0]{\spacefactor3000\relax}%
\providecommand \BibitemShut  [1]{\csname bibitem#1\endcsname}%
\let\auto@bib@innerbib\@empty
\bibitem [{\citenamefont {Geim}\ and\ \citenamefont
  {Grigorieva}(2013)}]{Geim2013}%
  \BibitemOpen
  \bibfield  {author} {\bibinfo {author} {\bibfnamefont {A.~K.}\ \bibnamefont
  {Geim}}\ and\ \bibinfo {author} {\bibfnamefont {I.~V.}\ \bibnamefont
  {Grigorieva}},\ }\bibfield  {title} {\bibinfo {title} {{Van der Waals
  heterostructures}},\ }\href {https://doi.org/10.1038/nature12385} {\bibfield
  {journal} {\bibinfo  {journal} {Nature}\ }\textbf {\bibinfo {volume} {499}},\
  \bibinfo {pages} {419} (\bibinfo {year} {2013})}\BibitemShut {NoStop}%
\bibitem [{\citenamefont {Arora}(2021)}]{Arora2021}%
  \BibitemOpen
  \bibfield  {author} {\bibinfo {author} {\bibfnamefont {A.}~\bibnamefont
  {Arora}},\ }\bibfield  {title} {\bibinfo {title} {{Magneto-optics of layered
  two-dimensional semiconductors and heterostructures: Progress and
  prospects}},\ }\href {https://doi.org/10.1063/5.0042683} {\bibfield
  {journal} {\bibinfo  {journal} {J. Appl. Phys.}\ }\textbf {\bibinfo {volume}
  {129}},\ \bibinfo {pages} {120902} (\bibinfo {year} {2021})}\BibitemShut
  {NoStop}%
\bibitem [{\citenamefont {Thi-Xuan~Dang}\ \emph {et~al.}(2022)\citenamefont
  {Thi-Xuan~Dang}, \citenamefont {Barik}, \citenamefont {Phan},\ and\
  \citenamefont {Woods}}]{Woods2022}%
  \BibitemOpen
  \bibfield  {author} {\bibinfo {author} {\bibfnamefont {D.}~\bibnamefont
  {Thi-Xuan~Dang}}, \bibinfo {author} {\bibfnamefont {R.~K.}\ \bibnamefont
  {Barik}}, \bibinfo {author} {\bibfnamefont {M.-H.}\ \bibnamefont {Phan}},\
  and\ \bibinfo {author} {\bibfnamefont {L.~M.}\ \bibnamefont {Woods}},\
  }\bibfield  {title} {\bibinfo {title} {{Enhanced magnetism in
  heterostructures with transition-metal dichalcogenide monolayers}},\ }\href
  {https://doi.org/10.1021/acs.jpclett.2c01925} {\bibfield  {journal} {\bibinfo
   {journal} {J. Phys. Chem. Lett.}\ }\textbf {\bibinfo {volume} {13}},\
  \bibinfo {pages} {8879} (\bibinfo {year} {2022})}\BibitemShut {NoStop}%
\bibitem [{\citenamefont {Phan}\ \emph {et~al.}(2023)\citenamefont {Phan},
  \citenamefont {Kalappattil}, \citenamefont {Jimenez}, \citenamefont {{Thi Hai
  Pham}}, \citenamefont {Mudiyanselage}, \citenamefont {Detellem},
  \citenamefont {Hung}, \citenamefont {Chanda},\ and\ \citenamefont
  {Eggers}}]{Phan2023}%
  \BibitemOpen
  \bibfield  {author} {\bibinfo {author} {\bibfnamefont {M.-H.}\ \bibnamefont
  {Phan}}, \bibinfo {author} {\bibfnamefont {V.}~\bibnamefont {Kalappattil}},
  \bibinfo {author} {\bibfnamefont {V.~O.}\ \bibnamefont {Jimenez}}, \bibinfo
  {author} {\bibfnamefont {Y.}~\bibnamefont {{Thi Hai Pham}}}, \bibinfo
  {author} {\bibfnamefont {N.~W.}\ \bibnamefont {Mudiyanselage}}, \bibinfo
  {author} {\bibfnamefont {D.}~\bibnamefont {Detellem}}, \bibinfo {author}
  {\bibfnamefont {C.-M.}\ \bibnamefont {Hung}}, \bibinfo {author}
  {\bibfnamefont {A.}~\bibnamefont {Chanda}},\ and\ \bibinfo {author}
  {\bibfnamefont {T.}~\bibnamefont {Eggers}},\ }\bibfield  {title} {\bibinfo
  {title} {{Exchange bias and interface-related effects in two-dimensional van
  der Waals magnetic heterostructures: Open questions and perspectives}},\
  }\href {https://doi.org/https://doi.org/10.1016/j.jallcom.2022.168375}
  {\bibfield  {journal} {\bibinfo  {journal} {J. Alloys and Compd.}\ }\textbf
  {\bibinfo {volume} {937}},\ \bibinfo {pages} {168375} (\bibinfo {year}
  {2023})}\BibitemShut {NoStop}%
\bibitem [{\citenamefont {Basov}\ \emph {et~al.}(2014)\citenamefont {Basov},
  \citenamefont {Fogler}, \citenamefont {Lanzara}, \citenamefont {Wang},\ and\
  \citenamefont {Zhang}}]{Basov2014}%
  \BibitemOpen
  \bibfield  {author} {\bibinfo {author} {\bibfnamefont {D.~N.}\ \bibnamefont
  {Basov}}, \bibinfo {author} {\bibfnamefont {M.~M.}\ \bibnamefont {Fogler}},
  \bibinfo {author} {\bibfnamefont {A.}~\bibnamefont {Lanzara}}, \bibinfo
  {author} {\bibfnamefont {F.}~\bibnamefont {Wang}},\ and\ \bibinfo {author}
  {\bibfnamefont {Y.}~\bibnamefont {Zhang}},\ }\bibfield  {title} {\bibinfo
  {title} {{Colloquium: Graphene spectroscopy}},\ }\href
  {https://doi.org/10.1103/RevModPhys.86.959} {\bibfield  {journal} {\bibinfo
  {journal} {Rev. Mod. Phys.}\ }\textbf {\bibinfo {volume} {86}},\ \bibinfo
  {pages} {959} (\bibinfo {year} {2014})}\BibitemShut {NoStop}%
\bibitem [{\citenamefont {Bussolotti}\ \emph {et~al.}(2021)\citenamefont
  {Bussolotti}, \citenamefont {Yang}, \citenamefont {Kawai}, \citenamefont
  {Chee},\ and\ \citenamefont {Goh}}]{Bussolotti2021}%
  \BibitemOpen
  \bibfield  {author} {\bibinfo {author} {\bibfnamefont {F.}~\bibnamefont
  {Bussolotti}}, \bibinfo {author} {\bibfnamefont {J.}~\bibnamefont {Yang}},
  \bibinfo {author} {\bibfnamefont {H.}~\bibnamefont {Kawai}}, \bibinfo
  {author} {\bibfnamefont {J.~Y.}\ \bibnamefont {Chee}},\ and\ \bibinfo
  {author} {\bibfnamefont {K.~E.~J.}\ \bibnamefont {Goh}},\ }\bibfield  {title}
  {\bibinfo {title} {{Influence of many-body effects on hole quasiparticle
  dynamics in a ${\mathrm{WS}}_{2}$ monolayer}},\ }\href
  {https://doi.org/10.1103/PhysRevB.103.045412} {\bibfield  {journal} {\bibinfo
   {journal} {Phys. Rev. B}\ }\textbf {\bibinfo {volume} {103}},\ \bibinfo
  {pages} {045412} (\bibinfo {year} {2021})}\BibitemShut {NoStop}%
\bibitem [{\citenamefont {Lee}\ \emph {et~al.}(2021)\citenamefont {Lee},
  \citenamefont {Lin}, \citenamefont {Lu}, \citenamefont {Chueh}, \citenamefont
  {Liu}, \citenamefont {Li}, \citenamefont {Chang}, \citenamefont {Kaindl},\
  and\ \citenamefont {Shih}}]{Lee2021}%
  \BibitemOpen
  \bibfield  {author} {\bibinfo {author} {\bibfnamefont {W.}~\bibnamefont
  {Lee}}, \bibinfo {author} {\bibfnamefont {Y.}~\bibnamefont {Lin}}, \bibinfo
  {author} {\bibfnamefont {L.-S.}\ \bibnamefont {Lu}}, \bibinfo {author}
  {\bibfnamefont {W.-C.}\ \bibnamefont {Chueh}}, \bibinfo {author}
  {\bibfnamefont {M.}~\bibnamefont {Liu}}, \bibinfo {author} {\bibfnamefont
  {X.}~\bibnamefont {Li}}, \bibinfo {author} {\bibfnamefont {W.-H.}\
  \bibnamefont {Chang}}, \bibinfo {author} {\bibfnamefont {R.~A.}\ \bibnamefont
  {Kaindl}},\ and\ \bibinfo {author} {\bibfnamefont {C.-K.}\ \bibnamefont
  {Shih}},\ }\bibfield  {title} {\bibinfo {title} {{Time-resolved ARPES
  determination of a quasi-particle band gap and hot electron dynamics in
  monolayer MoS2}},\ }\href {https://doi.org/10.1021/acs.nanolett.1c02674}
  {\bibfield  {journal} {\bibinfo  {journal} {Nano Lett.}\ }\textbf {\bibinfo
  {volume} {21}},\ \bibinfo {pages} {7363} (\bibinfo {year}
  {2021})}\BibitemShut {NoStop}%
\bibitem [{\citenamefont {Lin}\ \emph {et~al.}(2022)\citenamefont {Lin},
  \citenamefont {Chan}, \citenamefont {Lee}, \citenamefont {Lu}, \citenamefont
  {Li}, \citenamefont {Chang}, \citenamefont {Shih}, \citenamefont {Kaindl},
  \citenamefont {Louie},\ and\ \citenamefont {Lanzara}}]{Lin2022}%
  \BibitemOpen
  \bibfield  {author} {\bibinfo {author} {\bibfnamefont {Y.}~\bibnamefont
  {Lin}}, \bibinfo {author} {\bibfnamefont {Y.-h.}\ \bibnamefont {Chan}},
  \bibinfo {author} {\bibfnamefont {W.}~\bibnamefont {Lee}}, \bibinfo {author}
  {\bibfnamefont {L.-S.}\ \bibnamefont {Lu}}, \bibinfo {author} {\bibfnamefont
  {Z.}~\bibnamefont {Li}}, \bibinfo {author} {\bibfnamefont {W.-H.}\
  \bibnamefont {Chang}}, \bibinfo {author} {\bibfnamefont {C.-K.}\ \bibnamefont
  {Shih}}, \bibinfo {author} {\bibfnamefont {R.~A.}\ \bibnamefont {Kaindl}},
  \bibinfo {author} {\bibfnamefont {S.~G.}\ \bibnamefont {Louie}},\ and\
  \bibinfo {author} {\bibfnamefont {A.}~\bibnamefont {Lanzara}},\ }\bibfield
  {title} {\bibinfo {title} {{Exciton-driven renormalization of quasiparticle
  band structure in monolayer ${\mathrm{MoS}}_{2}$}},\ }\href
  {https://doi.org/10.1103/PhysRevB.106.L081117} {\bibfield  {journal}
  {\bibinfo  {journal} {Phys. Rev. B}\ }\textbf {\bibinfo {volume} {106}},\
  \bibinfo {pages} {L081117} (\bibinfo {year} {2022})}\BibitemShut {NoStop}%
\bibitem [{\citenamefont {Xiao}\ \emph {et~al.}(2012)\citenamefont {Xiao},
  \citenamefont {Liu}, \citenamefont {Feng}, \citenamefont {Xu},\ and\
  \citenamefont {Yao}}]{xiao2012}%
  \BibitemOpen
  \bibfield  {author} {\bibinfo {author} {\bibfnamefont {D.}~\bibnamefont
  {Xiao}}, \bibinfo {author} {\bibfnamefont {G.-B.}\ \bibnamefont {Liu}},
  \bibinfo {author} {\bibfnamefont {W.}~\bibnamefont {Feng}}, \bibinfo {author}
  {\bibfnamefont {X.}~\bibnamefont {Xu}},\ and\ \bibinfo {author}
  {\bibfnamefont {W.}~\bibnamefont {Yao}},\ }\bibfield  {title} {\bibinfo
  {title} {{Coupled spin and valley physics in monolayers of
  ${\mathrm{MoS}}_{2}$ and other group-VI dichalcogenides}},\ }\href
  {https://doi.org/10.1103/PhysRevLett.108.196802} {\bibfield  {journal}
  {\bibinfo  {journal} {Phys. Rev. Lett.}\ }\textbf {\bibinfo {volume} {108}},\
  \bibinfo {pages} {196802} (\bibinfo {year} {2012})}\BibitemShut {NoStop}%
\bibitem [{\citenamefont {Berkelbach}\ \emph {et~al.}(2013)\citenamefont
  {Berkelbach}, \citenamefont {Hybertsen},\ and\ \citenamefont
  {Reichman}}]{berkelbach2013}%
  \BibitemOpen
  \bibfield  {author} {\bibinfo {author} {\bibfnamefont {T.~C.}\ \bibnamefont
  {Berkelbach}}, \bibinfo {author} {\bibfnamefont {M.~S.}\ \bibnamefont
  {Hybertsen}},\ and\ \bibinfo {author} {\bibfnamefont {D.~R.}\ \bibnamefont
  {Reichman}},\ }\bibfield  {title} {\bibinfo {title} {{Theory of neutral and
  charged excitons in monolayer transition metal dichalcogenides}},\ }\href
  {https://doi.org/10.1103/PhysRevB.88.045318} {\bibfield  {journal} {\bibinfo
  {journal} {Phys. Rev. B}\ }\textbf {\bibinfo {volume} {88}},\ \bibinfo
  {pages} {045318} (\bibinfo {year} {2013})}\BibitemShut {NoStop}%
\bibitem [{\citenamefont {Korm{\'{a}}nyos}\ \emph {et~al.}(2015)\citenamefont
  {Korm{\'{a}}nyos}, \citenamefont {Burkard}, \citenamefont {Gmitra},
  \citenamefont {Fabian}, \citenamefont {Z{\'{o}}lyomi}, \citenamefont
  {Drummond},\ and\ \citenamefont {Fal'ko}}]{Korm2015}%
  \BibitemOpen
  \bibfield  {author} {\bibinfo {author} {\bibfnamefont {A.}~\bibnamefont
  {Korm{\'{a}}nyos}}, \bibinfo {author} {\bibfnamefont {G.}~\bibnamefont
  {Burkard}}, \bibinfo {author} {\bibfnamefont {M.}~\bibnamefont {Gmitra}},
  \bibinfo {author} {\bibfnamefont {J.}~\bibnamefont {Fabian}}, \bibinfo
  {author} {\bibfnamefont {V.}~\bibnamefont {Z{\'{o}}lyomi}}, \bibinfo {author}
  {\bibfnamefont {N.~D.}\ \bibnamefont {Drummond}},\ and\ \bibinfo {author}
  {\bibfnamefont {V.}~\bibnamefont {Fal'ko}},\ }\bibfield  {title} {\bibinfo
  {title} {{k $\cdotp$ p theory for two-dimensional transition metal
  dichalcogenide semiconductors}},\ }\href
  {https://doi.org/10.1088/2053-1583/2/2/022001} {\bibfield  {journal}
  {\bibinfo  {journal} {2D Materials}\ }\textbf {\bibinfo {volume} {2}},\
  \bibinfo {pages} {022001} (\bibinfo {year} {2015})}\BibitemShut {NoStop}%
\bibitem [{\citenamefont {Stier}\ \emph {et~al.}(2018)\citenamefont {Stier},
  \citenamefont {Wilson}, \citenamefont {Velizhanin}, \citenamefont {Kono},
  \citenamefont {Xu},\ and\ \citenamefont {Crooker}}]{PRL2018}%
  \BibitemOpen
  \bibfield  {author} {\bibinfo {author} {\bibfnamefont {A.~V.}\ \bibnamefont
  {Stier}}, \bibinfo {author} {\bibfnamefont {N.~P.}\ \bibnamefont {Wilson}},
  \bibinfo {author} {\bibfnamefont {K.~A.}\ \bibnamefont {Velizhanin}},
  \bibinfo {author} {\bibfnamefont {J.}~\bibnamefont {Kono}}, \bibinfo {author}
  {\bibfnamefont {X.}~\bibnamefont {Xu}},\ and\ \bibinfo {author}
  {\bibfnamefont {S.~A.}\ \bibnamefont {Crooker}},\ }\bibfield  {title}
  {\bibinfo {title} {Magnetooptics of exciton {Rydberg} states in a monolayer
  semiconductor},\ }\href {https://doi.org/10.1103/PhysRevLett.120.057405}
  {\bibfield  {journal} {\bibinfo  {journal} {Phys. Rev. Lett.}\ }\textbf
  {\bibinfo {volume} {120}},\ \bibinfo {pages} {057405} (\bibinfo {year}
  {2018})}\BibitemShut {NoStop}%
\bibitem [{\citenamefont {Goryca}\ \emph {et~al.}(2019)\citenamefont {Goryca},
  \citenamefont {Li}, \citenamefont {Stier}, \citenamefont {Taniguchi},
  \citenamefont {Watanabe}, \citenamefont {Courtade}, \citenamefont {Shree},
  \citenamefont {Robert}, \citenamefont {Urbaszek}, \citenamefont {Marie},\
  and\ \citenamefont {Crooker}}]{NAT2019}%
  \BibitemOpen
  \bibfield  {author} {\bibinfo {author} {\bibfnamefont {M.}~\bibnamefont
  {Goryca}}, \bibinfo {author} {\bibfnamefont {J.}~\bibnamefont {Li}}, \bibinfo
  {author} {\bibfnamefont {A.~V.}\ \bibnamefont {Stier}}, \bibinfo {author}
  {\bibfnamefont {T.}~\bibnamefont {Taniguchi}}, \bibinfo {author}
  {\bibfnamefont {K.}~\bibnamefont {Watanabe}}, \bibinfo {author}
  {\bibfnamefont {E.}~\bibnamefont {Courtade}}, \bibinfo {author}
  {\bibfnamefont {S.}~\bibnamefont {Shree}}, \bibinfo {author} {\bibfnamefont
  {C.}~\bibnamefont {Robert}}, \bibinfo {author} {\bibfnamefont
  {B.}~\bibnamefont {Urbaszek}}, \bibinfo {author} {\bibfnamefont
  {X.}~\bibnamefont {Marie}},\ and\ \bibinfo {author} {\bibfnamefont {S.~A.}\
  \bibnamefont {Crooker}},\ }\bibfield  {title} {\bibinfo {title} {{Revealing
  exciton masses and dielectric properties of monolayer semiconductors with
  high magnetic fields}},\ }\href {https://doi.org/10.1038/s41467-019-12180-y}
  {\bibfield  {journal} {\bibinfo  {journal} {Nat. Commun.}\ }\textbf {\bibinfo
  {volume} {10}},\ \bibinfo {pages} {4172} (\bibinfo {year}
  {2019})}\BibitemShut {NoStop}%
\bibitem [{\citenamefont {Liu}\ \emph {et~al.}(2019)\citenamefont {Liu},
  \citenamefont {van Baren}, \citenamefont {Taniguchi}, \citenamefont
  {Watanabe}, \citenamefont {Chang},\ and\ \citenamefont {Lui}}]{Liu2019}%
  \BibitemOpen
  \bibfield  {author} {\bibinfo {author} {\bibfnamefont {E.}~\bibnamefont
  {Liu}}, \bibinfo {author} {\bibfnamefont {J.}~\bibnamefont {van Baren}},
  \bibinfo {author} {\bibfnamefont {T.}~\bibnamefont {Taniguchi}}, \bibinfo
  {author} {\bibfnamefont {K.}~\bibnamefont {Watanabe}}, \bibinfo {author}
  {\bibfnamefont {Y.-C.}\ \bibnamefont {Chang}},\ and\ \bibinfo {author}
  {\bibfnamefont {C.~H.}\ \bibnamefont {Lui}},\ }\bibfield  {title} {\bibinfo
  {title} {Magnetophotoluminescence of exciton rydberg states in monolayer
  $\mathrm{WS}{\mathrm{e}}_{2}$},\ }\href
  {https://doi.org/10.1103/PhysRevB.99.205420} {\bibfield  {journal} {\bibinfo
  {journal} {Phys. Rev. B}\ }\textbf {\bibinfo {volume} {99}},\ \bibinfo
  {pages} {205420} (\bibinfo {year} {2019})}\BibitemShut {NoStop}%
\bibitem [{\citenamefont {Chernikov}\ \emph {et~al.}(2014)\citenamefont
  {Chernikov}, \citenamefont {Berkelbach}, \citenamefont {Hill}, \citenamefont
  {Rigosi}, \citenamefont {Li}, \citenamefont {Aslan}, \citenamefont
  {Reichman}, \citenamefont {Hybertsen},\ and\ \citenamefont
  {Heinz}}]{chernikov2014}%
  \BibitemOpen
  \bibfield  {author} {\bibinfo {author} {\bibfnamefont {A.}~\bibnamefont
  {Chernikov}}, \bibinfo {author} {\bibfnamefont {T.~C.}\ \bibnamefont
  {Berkelbach}}, \bibinfo {author} {\bibfnamefont {H.~M.}\ \bibnamefont
  {Hill}}, \bibinfo {author} {\bibfnamefont {A.}~\bibnamefont {Rigosi}},
  \bibinfo {author} {\bibfnamefont {Y.}~\bibnamefont {Li}}, \bibinfo {author}
  {\bibfnamefont {O.~B.}\ \bibnamefont {Aslan}}, \bibinfo {author}
  {\bibfnamefont {D.~R.}\ \bibnamefont {Reichman}}, \bibinfo {author}
  {\bibfnamefont {M.~S.}\ \bibnamefont {Hybertsen}},\ and\ \bibinfo {author}
  {\bibfnamefont {T.~F.}\ \bibnamefont {Heinz}},\ }\bibfield  {title} {\bibinfo
  {title} {Exciton binding energy and nonhydrogenic {Rydberg} series in
  monolayer {${\mathrm{WS}}_{2}$}},\ }\href
  {https://doi.org/10.1103/PhysRevLett.113.076802} {\bibfield  {journal}
  {\bibinfo  {journal} {Phys. Rev. Lett.}\ }\textbf {\bibinfo {volume} {113}},\
  \bibinfo {pages} {076802} (\bibinfo {year} {2014})}\BibitemShut {NoStop}%
\bibitem [{\citenamefont {Nguyen}\ \emph {et~al.}(2019)\citenamefont {Nguyen},
  \citenamefont {Ly}, \citenamefont {Le}, \citenamefont {Hoang},\ and\
  \citenamefont {Le}}]{PhysE}%
  \BibitemOpen
  \bibfield  {author} {\bibinfo {author} {\bibfnamefont {D.-A.~P.}\
  \bibnamefont {Nguyen}}, \bibinfo {author} {\bibfnamefont {D.-N.}\
  \bibnamefont {Ly}}, \bibinfo {author} {\bibfnamefont {D.-N.}\ \bibnamefont
  {Le}}, \bibinfo {author} {\bibfnamefont {N.-T.~D.}\ \bibnamefont {Hoang}},\
  and\ \bibinfo {author} {\bibfnamefont {V.-H.}\ \bibnamefont {Le}},\
  }\bibfield  {title} {\bibinfo {title} {"high-accuracy energy spectra of a
  two-dimensional exciton screened by reduced dimensionality with the presence
  of a constant magnetic field"},\ }\href
  {https://doi.org/https://doi.org/10.1016/j.physe.2019.05.007} {\bibfield
  {journal} {\bibinfo  {journal} {Physica E}\ }\textbf {\bibinfo {volume}
  {113}},\ \bibinfo {pages} {152 } (\bibinfo {year} {2019})}\BibitemShut
  {NoStop}%
\bibitem [{\citenamefont {Plechinger}\ \emph {et~al.}(2016)\citenamefont
  {Plechinger}, \citenamefont {Nagler}, \citenamefont {Arora}, \citenamefont
  {Granados~del Águila}, \citenamefont {Ballottin}, \citenamefont {Frank},
  \citenamefont {Steinleitner}, \citenamefont {Gmitra}, \citenamefont {Fabian},
  \citenamefont {Christianen}, \citenamefont {Bratschitsch}, \citenamefont
  {Schüller},\ and\ \citenamefont {Korn}}]{Plechinger-2016}%
  \BibitemOpen
  \bibfield  {author} {\bibinfo {author} {\bibfnamefont {G.}~\bibnamefont
  {Plechinger}}, \bibinfo {author} {\bibfnamefont {P.}~\bibnamefont {Nagler}},
  \bibinfo {author} {\bibfnamefont {A.}~\bibnamefont {Arora}}, \bibinfo
  {author} {\bibfnamefont {A.}~\bibnamefont {Granados~del Águila}}, \bibinfo
  {author} {\bibfnamefont {M.~V.}\ \bibnamefont {Ballottin}}, \bibinfo {author}
  {\bibfnamefont {T.}~\bibnamefont {Frank}}, \bibinfo {author} {\bibfnamefont
  {P.}~\bibnamefont {Steinleitner}}, \bibinfo {author} {\bibfnamefont
  {M.}~\bibnamefont {Gmitra}}, \bibinfo {author} {\bibfnamefont
  {J.}~\bibnamefont {Fabian}}, \bibinfo {author} {\bibfnamefont {P.~C.~M.}\
  \bibnamefont {Christianen}}, \bibinfo {author} {\bibfnamefont
  {R.}~\bibnamefont {Bratschitsch}}, \bibinfo {author} {\bibfnamefont
  {C.}~\bibnamefont {Schüller}},\ and\ \bibinfo {author} {\bibfnamefont
  {T.}~\bibnamefont {Korn}},\ }\bibfield  {title} {\bibinfo {title} {Excitonic
  valley effects in monolayer ws2 under high magnetic fields},\ }\href
  {https://doi.org/10.1021/acs.nanolett.6b04171} {\bibfield  {journal}
  {\bibinfo  {journal} {Nano Lett.}\ }\textbf {\bibinfo {volume} {16}},\
  \bibinfo {pages} {7899} (\bibinfo {year} {2016})}\BibitemShut {NoStop}%
\bibitem [{\citenamefont {Stier}\ \emph
  {et~al.}(2016{\natexlab{a}})\citenamefont {Stier}, \citenamefont {McCreary},
  \citenamefont {Jonker}, \citenamefont {Kono},\ and\ \citenamefont
  {Crooker}}]{Stier2016-nat}%
  \BibitemOpen
  \bibfield  {author} {\bibinfo {author} {\bibfnamefont {A.~V.}\ \bibnamefont
  {Stier}}, \bibinfo {author} {\bibfnamefont {K.~M.}\ \bibnamefont {McCreary}},
  \bibinfo {author} {\bibfnamefont {B.~T.}\ \bibnamefont {Jonker}}, \bibinfo
  {author} {\bibfnamefont {J.}~\bibnamefont {Kono}},\ and\ \bibinfo {author}
  {\bibfnamefont {S.~A.}\ \bibnamefont {Crooker}},\ }\bibfield  {title}
  {\bibinfo {title} {Exciton diamagnetic shifts and valley zeeman effects in
  monolayer {WS$_2$} and {MoS$_2$} to {$65$ Tesla}},\ }\href
  {https://doi.org/10.1038/ncomms10643} {\bibfield  {journal} {\bibinfo
  {journal} {Nat. Commun.}\ }\textbf {\bibinfo {volume} {7}},\ \bibinfo {pages}
  {10643} (\bibinfo {year} {2016}{\natexlab{a}})}\BibitemShut {NoStop}%
\bibitem [{\citenamefont {Stier}\ \emph
  {et~al.}(2016{\natexlab{b}})\citenamefont {Stier}, \citenamefont {Wilson},
  \citenamefont {Clark}, \citenamefont {Xu},\ and\ \citenamefont
  {Crooker}}]{Stier2016-nano}%
  \BibitemOpen
  \bibfield  {author} {\bibinfo {author} {\bibfnamefont {A.~V.}\ \bibnamefont
  {Stier}}, \bibinfo {author} {\bibfnamefont {N.~P.}\ \bibnamefont {Wilson}},
  \bibinfo {author} {\bibfnamefont {G.}~\bibnamefont {Clark}}, \bibinfo
  {author} {\bibfnamefont {X.}~\bibnamefont {Xu}},\ and\ \bibinfo {author}
  {\bibfnamefont {S.~A.}\ \bibnamefont {Crooker}},\ }\bibfield  {title}
  {\bibinfo {title} {Probing the influence of dielectric environment on
  excitons in monolayer {${\mathrm{WSe}}_{2}$}: Insight from high magnetic
  fields},\ }\href {https://doi.org/10.1021/acs.nanolett.6b03276} {\bibfield
  {journal} {\bibinfo  {journal} {Nano Lett.}\ }\textbf {\bibinfo {volume}
  {16}},\ \bibinfo {pages} {7054} (\bibinfo {year}
  {2016}{\natexlab{b}})}\BibitemShut {NoStop}%
\bibitem [{\citenamefont {Zipfel}\ \emph {et~al.}(2018)\citenamefont {Zipfel},
  \citenamefont {Holler}, \citenamefont {Mitioglu}, \citenamefont {Ballottin},
  \citenamefont {Nagler}, \citenamefont {Stier}, \citenamefont {Taniguchi},
  \citenamefont {Watanabe}, \citenamefont {Crooker}, \citenamefont
  {Christianen}, \citenamefont {Korn},\ and\ \citenamefont
  {Chernikov}}]{Zipfel2018}%
  \BibitemOpen
  \bibfield  {author} {\bibinfo {author} {\bibfnamefont {J.}~\bibnamefont
  {Zipfel}}, \bibinfo {author} {\bibfnamefont {J.}~\bibnamefont {Holler}},
  \bibinfo {author} {\bibfnamefont {A.~A.}\ \bibnamefont {Mitioglu}}, \bibinfo
  {author} {\bibfnamefont {M.~V.}\ \bibnamefont {Ballottin}}, \bibinfo {author}
  {\bibfnamefont {P.}~\bibnamefont {Nagler}}, \bibinfo {author} {\bibfnamefont
  {A.~V.}\ \bibnamefont {Stier}}, \bibinfo {author} {\bibfnamefont
  {T.}~\bibnamefont {Taniguchi}}, \bibinfo {author} {\bibfnamefont
  {K.}~\bibnamefont {Watanabe}}, \bibinfo {author} {\bibfnamefont {S.~A.}\
  \bibnamefont {Crooker}}, \bibinfo {author} {\bibfnamefont {P.~C.~M.}\
  \bibnamefont {Christianen}}, \bibinfo {author} {\bibfnamefont
  {T.}~\bibnamefont {Korn}},\ and\ \bibinfo {author} {\bibfnamefont
  {A.}~\bibnamefont {Chernikov}},\ }\bibfield  {title} {\bibinfo {title}
  {Spatial extent of the excited exciton states in ${\mathrm{ws}}_{2}$
  monolayers from diamagnetic shifts},\ }\href
  {https://doi.org/10.1103/PhysRevB.98.075438} {\bibfield  {journal} {\bibinfo
  {journal} {Phys. Rev. B}\ }\textbf {\bibinfo {volume} {98}},\ \bibinfo
  {pages} {075438} (\bibinfo {year} {2018})}\BibitemShut {NoStop}%
\bibitem [{\citenamefont {Chen}\ \emph {et~al.}(2019)\citenamefont {Chen},
  \citenamefont {Lu}, \citenamefont {Goldstein}, \citenamefont {Tong},
  \citenamefont {Chaves}, \citenamefont {Kunstmann}, \citenamefont
  {Cavalcante}, \citenamefont {Woźniak}, \citenamefont {Seifert},
  \citenamefont {Reichman}, \citenamefont {Taniguchi}, \citenamefont
  {Watanabe}, \citenamefont {Smirnov},\ and\ \citenamefont
  {Yan}}]{Chen2019-nano}%
  \BibitemOpen
  \bibfield  {author} {\bibinfo {author} {\bibfnamefont {S.-Y.}\ \bibnamefont
  {Chen}}, \bibinfo {author} {\bibfnamefont {Z.}~\bibnamefont {Lu}}, \bibinfo
  {author} {\bibfnamefont {T.}~\bibnamefont {Goldstein}}, \bibinfo {author}
  {\bibfnamefont {J.}~\bibnamefont {Tong}}, \bibinfo {author} {\bibfnamefont
  {A.}~\bibnamefont {Chaves}}, \bibinfo {author} {\bibfnamefont
  {J.}~\bibnamefont {Kunstmann}}, \bibinfo {author} {\bibfnamefont {L.~S.~R.}\
  \bibnamefont {Cavalcante}}, \bibinfo {author} {\bibfnamefont
  {T.}~\bibnamefont {Woźniak}}, \bibinfo {author} {\bibfnamefont
  {G.}~\bibnamefont {Seifert}}, \bibinfo {author} {\bibfnamefont {D.~R.}\
  \bibnamefont {Reichman}}, \bibinfo {author} {\bibfnamefont {T.}~\bibnamefont
  {Taniguchi}}, \bibinfo {author} {\bibfnamefont {K.}~\bibnamefont {Watanabe}},
  \bibinfo {author} {\bibfnamefont {D.}~\bibnamefont {Smirnov}},\ and\ \bibinfo
  {author} {\bibfnamefont {J.}~\bibnamefont {Yan}},\ }\bibfield  {title}
  {\bibinfo {title} {Luminescent emission of excited rydberg excitons from
  monolayer {WSe$_2$}},\ }\href {https://doi.org/10.1021/acs.nanolett.9b00029}
  {\bibfield  {journal} {\bibinfo  {journal} {Nano Lett.}\ }\textbf {\bibinfo
  {volume} {19}},\ \bibinfo {pages} {2464} (\bibinfo {year}
  {2019})}\BibitemShut {NoStop}%
\bibitem [{\citenamefont {Rytova}(1967)}]{Rytova1967}%
  \BibitemOpen
  \bibfield  {author} {\bibinfo {author} {\bibfnamefont {N.~S.}\ \bibnamefont
  {Rytova}},\ }\bibfield  {title} {\bibinfo {title} {{Screened potential of a
  point charge in a thin film}},\ }\href@noop {} {\bibfield  {journal}
  {\bibinfo  {journal} {Moscow University Physics Bulletin}\ }\textbf {\bibinfo
  {volume} {22}},\ \bibinfo {pages} {30} (\bibinfo {year} {1967})}\BibitemShut
  {NoStop}%
\bibitem [{\citenamefont {Keldysh}(1979)}]{keldysh1979}%
  \BibitemOpen
  \bibfield  {author} {\bibinfo {author} {\bibfnamefont {L.~V.}\ \bibnamefont
  {Keldysh}},\ }\bibfield  {title} {\bibinfo {title} {Coulomb interaction in
  thin semiconductor and semimetal films},\ }\href
  {http://www.jetpletters.ac.ru/ps/1458/article-22207.shtml} {\bibfield
  {journal} {\bibinfo  {journal} {JETP Lett.}\ }\textbf {\bibinfo {volume}
  {29}},\ \bibinfo {pages} {658} (\bibinfo {year} {1979})}\BibitemShut
  {NoStop}%
\bibitem [{\citenamefont {Hanamura}\ \emph {et~al.}(1988)\citenamefont
  {Hanamura}, \citenamefont {Nagaosa}, \citenamefont {Kumagai},\ and\
  \citenamefont {Takagahara}}]{haramura1988}%
  \BibitemOpen
  \bibfield  {author} {\bibinfo {author} {\bibfnamefont {E.}~\bibnamefont
  {Hanamura}}, \bibinfo {author} {\bibfnamefont {N.}~\bibnamefont {Nagaosa}},
  \bibinfo {author} {\bibfnamefont {M.}~\bibnamefont {Kumagai}},\ and\ \bibinfo
  {author} {\bibfnamefont {T.}~\bibnamefont {Takagahara}},\ }\bibfield  {title}
  {\bibinfo {title} {Quantum wells with enhanced exciton effects and optical
  non-linearity},\ }\href {https://doi.org/10.1016/0921-5107(88)90006-2}
  {\bibfield  {journal} {\bibinfo  {journal} {Mater. Sci. Eng. B: Solid-State
  Mater. Adv. Technol.}\ }\textbf {\bibinfo {volume} {1}},\ \bibinfo {pages}
  {255} (\bibinfo {year} {1988})}\BibitemShut {NoStop}%
\bibitem [{\citenamefont {Cudazzo}\ \emph {et~al.}(2011)\citenamefont
  {Cudazzo}, \citenamefont {Tokatly},\ and\ \citenamefont
  {Rubio}}]{cudazzo2011}%
  \BibitemOpen
  \bibfield  {author} {\bibinfo {author} {\bibfnamefont {P.}~\bibnamefont
  {Cudazzo}}, \bibinfo {author} {\bibfnamefont {I.~V.}\ \bibnamefont
  {Tokatly}},\ and\ \bibinfo {author} {\bibfnamefont {A.}~\bibnamefont
  {Rubio}},\ }\bibfield  {title} {\bibinfo {title} {{Dielectric screening in
  two-dimensional insulators: Implications for excitonic and impurity states in
  graphene}},\ }\href {https://doi.org/10.1103/PhysRevB.84.085406} {\bibfield
  {journal} {\bibinfo  {journal} {Phys. Rev. B}\ }\textbf {\bibinfo {volume}
  {84}},\ \bibinfo {pages} {085406} (\bibinfo {year} {2011})}\BibitemShut
  {NoStop}%
\bibitem [{\citenamefont {Molas}\ \emph {et~al.}(2019)\citenamefont {Molas},
  \citenamefont {Slobodeniuk}, \citenamefont {Nogajewski}, \citenamefont
  {Bartos}, \citenamefont {Bala}, \citenamefont {Babi\ifmmode~\acute{n}\else
  \'{n}\fi{}ski}, \citenamefont {Watanabe}, \citenamefont {Taniguchi},
  \citenamefont {Faugeras},\ and\ \citenamefont {Potemski}}]{Molas2019}%
  \BibitemOpen
  \bibfield  {author} {\bibinfo {author} {\bibfnamefont {M.~R.}\ \bibnamefont
  {Molas}}, \bibinfo {author} {\bibfnamefont {A.~O.}\ \bibnamefont
  {Slobodeniuk}}, \bibinfo {author} {\bibfnamefont {K.}~\bibnamefont
  {Nogajewski}}, \bibinfo {author} {\bibfnamefont {M.}~\bibnamefont {Bartos}},
  \bibinfo {author} {\bibfnamefont {L.}~\bibnamefont {Bala}}, \bibinfo {author}
  {\bibfnamefont {A.}~\bibnamefont {Babi\ifmmode~\acute{n}\else
  \'{n}\fi{}ski}}, \bibinfo {author} {\bibfnamefont {K.}~\bibnamefont
  {Watanabe}}, \bibinfo {author} {\bibfnamefont {T.}~\bibnamefont {Taniguchi}},
  \bibinfo {author} {\bibfnamefont {C.}~\bibnamefont {Faugeras}},\ and\
  \bibinfo {author} {\bibfnamefont {M.}~\bibnamefont {Potemski}},\ }\bibfield
  {title} {\bibinfo {title} {Energy spectrum of two-dimensional excitons in a
  nonuniform dielectric medium},\ }\href
  {https://doi.org/10.1103/PhysRevLett.123.136801} {\bibfield  {journal}
  {\bibinfo  {journal} {Phys. Rev. Lett.}\ }\textbf {\bibinfo {volume} {123}},\
  \bibinfo {pages} {136801} (\bibinfo {year} {2019})}\BibitemShut {NoStop}%
\bibitem [{\citenamefont {Nguyen-Truong}(2022)}]{Hieu2022}%
  \BibitemOpen
  \bibfield  {author} {\bibinfo {author} {\bibfnamefont {H.~T.}\ \bibnamefont
  {Nguyen-Truong}},\ }\bibfield  {title} {\bibinfo {title} {Exciton binding
  energy and screening length in two-dimensional semiconductors},\ }\href
  {https://doi.org/10.1103/PhysRevB.105.L201407} {\bibfield  {journal}
  {\bibinfo  {journal} {Phys. Rev. B}\ }\textbf {\bibinfo {volume} {105}},\
  \bibinfo {pages} {L201407} (\bibinfo {year} {2022})}\BibitemShut {NoStop}%
\bibitem [{\citenamefont {Feranchuk}\ and\ \citenamefont
  {Komarov}(1982)}]{Feranchuk1982}%
  \BibitemOpen
  \bibfield  {author} {\bibinfo {author} {\bibfnamefont {I.}~\bibnamefont
  {Feranchuk}}\ and\ \bibinfo {author} {\bibfnamefont {L.}~\bibnamefont
  {Komarov}},\ }\bibfield  {title} {\bibinfo {title} {The operator method of
  the approximate solution of the {S}chr{\"o}dinger equation},\ }\href
  {https://doi.org/10.1016/0375-9601(82)90229-8} {\bibfield  {journal}
  {\bibinfo  {journal} {Phys. Lett. A}\ }\textbf {\bibinfo {volume} {88}},\
  \bibinfo {pages} {211 } (\bibinfo {year} {1982})}\BibitemShut {NoStop}%
\bibitem [{\citenamefont {Feranchuk}\ \emph {et~al.}(2015)\citenamefont
  {Feranchuk}, \citenamefont {Ivanov}, \citenamefont {Le},\ and\ \citenamefont
  {Ulyanenkov}}]{Hoangbook2015}%
  \BibitemOpen
  \bibfield  {author} {\bibinfo {author} {\bibfnamefont {I.}~\bibnamefont
  {Feranchuk}}, \bibinfo {author} {\bibfnamefont {A.}~\bibnamefont {Ivanov}},
  \bibinfo {author} {\bibfnamefont {V.-H.}\ \bibnamefont {Le}},\ and\ \bibinfo
  {author} {\bibfnamefont {A.}~\bibnamefont {Ulyanenkov}},\ }\href
  {https://doi.org/10.1007/978-3-319-13006-4} {\emph {\bibinfo {title}
  {{Non-perturbative Description of Quantum Systems}}}}\ (\bibinfo  {publisher}
  {Springer},\ \bibinfo {address} {Switzerland},\ \bibinfo {year}
  {2015})\BibitemShut {NoStop}%
\bibitem [{\citenamefont {Hoang}\ \emph {et~al.}(2020)\citenamefont {Hoang},
  \citenamefont {Ly},\ and\ \citenamefont {Le}}]{comment2020}%
  \BibitemOpen
  \bibfield  {author} {\bibinfo {author} {\bibfnamefont {D.~N.-T.}\
  \bibnamefont {Hoang}}, \bibinfo {author} {\bibfnamefont {D.-N.}\ \bibnamefont
  {Ly}},\ and\ \bibinfo {author} {\bibfnamefont {V.-H.}\ \bibnamefont {Le}},\
  }\bibfield  {title} {\bibinfo {title} {Comment on \lq\lq{Excitons, trions,
  and biexcitons in transition-metal dichalcogenides: Magnetic-field
  dependence}\rq\rq{}},\ }\href {https://doi.org/10.1103/PhysRevB.101.127401}
  {\bibfield  {journal} {\bibinfo  {journal} {Phys. Rev. B}\ }\textbf {\bibinfo
  {volume} {115}},\ \bibinfo {pages} {127401} (\bibinfo {year}
  {2020})}\BibitemShut {NoStop}%
\bibitem [{\citenamefont {Ly}\ \emph {et~al.}(2023)\citenamefont {Ly},
  \citenamefont {Le}, \citenamefont {Phan},\ and\ \citenamefont
  {Le}}]{Nhat2022}%
  \BibitemOpen
  \bibfield  {author} {\bibinfo {author} {\bibfnamefont {D.-N.}\ \bibnamefont
  {Ly}}, \bibinfo {author} {\bibfnamefont {D.-N.}\ \bibnamefont {Le}}, \bibinfo
  {author} {\bibfnamefont {N.-H.}\ \bibnamefont {Phan}},\ and\ \bibinfo
  {author} {\bibfnamefont {V.-H.}\ \bibnamefont {Le}},\ }\bibfield  {title}
  {\bibinfo {title} {Thermal effect on magnetoexciton energy spectra in
  monolayer transition metal dichalcogenides},\ }\href
  {https://doi.org/10.1103/PhysRevB.107.155410} {\bibfield  {journal} {\bibinfo
   {journal} {Phys. Rev. B}\ }\textbf {\bibinfo {volume} {107}},\ \bibinfo
  {pages} {155410} (\bibinfo {year} {2023})}\BibitemShut {NoStop}%
\bibitem [{\citenamefont {Taghizadeh}\ and\ \citenamefont
  {Pedersen}(2019)}]{Taghizadeh2019}%
  \BibitemOpen
  \bibfield  {author} {\bibinfo {author} {\bibfnamefont {A.}~\bibnamefont
  {Taghizadeh}}\ and\ \bibinfo {author} {\bibfnamefont {T.~G.}\ \bibnamefont
  {Pedersen}},\ }\bibfield  {title} {\bibinfo {title} {Nonlinear optical
  selection rules of excitons in monolayer transition metal dichalcogenides},\
  }\href {https://doi.org/10.1103/PhysRevB.99.235433} {\bibfield  {journal}
  {\bibinfo  {journal} {Phys. Rev. B}\ }\textbf {\bibinfo {volume} {99}},\
  \bibinfo {pages} {235433} (\bibinfo {year} {2019})}\BibitemShut {NoStop}%
\bibitem [{\citenamefont {Henriques}\ \emph {et~al.}(2021)\citenamefont
  {Henriques}, \citenamefont {Kamban}, \citenamefont {Pedersen},\ and\
  \citenamefont {Peres}}]{Henriques2021}%
  \BibitemOpen
  \bibfield  {author} {\bibinfo {author} {\bibfnamefont {J.~C.~G.}\
  \bibnamefont {Henriques}}, \bibinfo {author} {\bibfnamefont {H.~C.}\
  \bibnamefont {Kamban}}, \bibinfo {author} {\bibfnamefont {T.~G.}\
  \bibnamefont {Pedersen}},\ and\ \bibinfo {author} {\bibfnamefont {N.~M.~R.}\
  \bibnamefont {Peres}},\ }\bibfield  {title} {\bibinfo {title} {Calculation of
  the nonlinear response functions of intraexciton transitions in
  two-dimensional transition metal dichalcogenides},\ }\href
  {https://doi.org/10.1103/PhysRevB.103.235412} {\bibfield  {journal} {\bibinfo
   {journal} {Phys. Rev. B}\ }\textbf {\bibinfo {volume} {103}},\ \bibinfo
  {pages} {235412} (\bibinfo {year} {2021})}\BibitemShut {NoStop}%
\bibitem [{\citenamefont {{Netlib.org. LAPACK: Linear Algebra
  PACKage}}()}]{Lapack}%
  \BibitemOpen
  \bibfield  {author} {\bibinfo {author} {\bibnamefont {{Netlib.org. LAPACK:
  Linear Algebra PACKage}}},\ }\href
  {{http://www.netlib.org/lapack/explore-ht-ml/d2/d97/dsyevx-8f.html}}
  {\bibinfo {title} {Subroutine dsygvx.f}}\BibitemShut {NoStop}%
\bibitem [{\citenamefont {Hoang}\ and\ \citenamefont
  {Giang}(1993)}]{giang1993}%
  \BibitemOpen
  \bibfield  {author} {\bibinfo {author} {\bibfnamefont {L.~V.}\ \bibnamefont
  {Hoang}}\ and\ \bibinfo {author} {\bibfnamefont {N.~T.}\ \bibnamefont
  {Giang}},\ }\bibfield  {title} {\bibinfo {title} {The algebraic method for
  two-dimensional quantum atomic systems},\ }\href
  {https://doi.org/10.1088/0305-4470/26/6/022} {\bibfield  {journal} {\bibinfo
  {journal} {J. Phys. A: Math. Gen.}\ }\textbf {\bibinfo {volume} {26}},\
  \bibinfo {pages} {1409} (\bibinfo {year} {1993})}\BibitemShut {NoStop}%
\bibitem [{\citenamefont {Haastrup}\ \emph {et~al.}(2018)\citenamefont
  {Haastrup}, \citenamefont {Strange}, \citenamefont {Pandey}, \citenamefont
  {Deilmann}, \citenamefont {Schmidt}, \citenamefont {Hinsche}, \citenamefont
  {Gjerding}, \citenamefont {Torelli}, \citenamefont {Larsen}, \citenamefont
  {Riis-Jensen}, \citenamefont {Gath}, \citenamefont {Jacobsen}, \citenamefont
  {Mortensen}, \citenamefont {Olsen},\ and\ \citenamefont
  {Thygesen}}]{haastrup2018}%
  \BibitemOpen
  \bibfield  {author} {\bibinfo {author} {\bibfnamefont {S.}~\bibnamefont
  {Haastrup}}, \bibinfo {author} {\bibfnamefont {M.}~\bibnamefont {Strange}},
  \bibinfo {author} {\bibfnamefont {M.}~\bibnamefont {Pandey}}, \bibinfo
  {author} {\bibfnamefont {T.}~\bibnamefont {Deilmann}}, \bibinfo {author}
  {\bibfnamefont {P.~S.}\ \bibnamefont {Schmidt}}, \bibinfo {author}
  {\bibfnamefont {N.~F.}\ \bibnamefont {Hinsche}}, \bibinfo {author}
  {\bibfnamefont {M.~N.}\ \bibnamefont {Gjerding}}, \bibinfo {author}
  {\bibfnamefont {D.}~\bibnamefont {Torelli}}, \bibinfo {author} {\bibfnamefont
  {P.~M.}\ \bibnamefont {Larsen}}, \bibinfo {author} {\bibfnamefont {A.~C.}\
  \bibnamefont {Riis-Jensen}}, \bibinfo {author} {\bibfnamefont
  {J.}~\bibnamefont {Gath}}, \bibinfo {author} {\bibfnamefont {K.~W.}\
  \bibnamefont {Jacobsen}}, \bibinfo {author} {\bibfnamefont {J.~J.}\
  \bibnamefont {Mortensen}}, \bibinfo {author} {\bibfnamefont {T.}~\bibnamefont
  {Olsen}},\ and\ \bibinfo {author} {\bibfnamefont {K.~S.}\ \bibnamefont
  {Thygesen}},\ }\bibfield  {title} {\bibinfo {title} {The computational 2d
  materials database: high-throughput modeling and discovery of atomically thin
  crystals},\ }\href {https://doi.org/10.1088/2053-1583/aacfc1} {\bibfield
  {journal} {\bibinfo  {journal} {2D Materials}\ }\textbf {\bibinfo {volume}
  {5}},\ \bibinfo {pages} {042002} (\bibinfo {year} {2018})}\BibitemShut
  {NoStop}%
\bibitem [{\citenamefont {Gjerding}\ \emph {et~al.}(2021)\citenamefont
  {Gjerding}, \citenamefont {Taghizadeh}, \citenamefont {Rasmussen},
  \citenamefont {Ali}, \citenamefont {Bertoldo}, \citenamefont {Deilmann},
  \citenamefont {Knøsgaard}, \citenamefont {Kruse}, \citenamefont {Larsen},
  \citenamefont {Manti}, \citenamefont {Pedersen}, \citenamefont {Petralanda},
  \citenamefont {Skovhus}, \citenamefont {Svendsen}, \citenamefont {Mortensen},
  \citenamefont {Olsen},\ and\ \citenamefont {Thygesen}}]{gjerding2021}%
  \BibitemOpen
  \bibfield  {author} {\bibinfo {author} {\bibfnamefont {M.~N.}\ \bibnamefont
  {Gjerding}}, \bibinfo {author} {\bibfnamefont {A.}~\bibnamefont
  {Taghizadeh}}, \bibinfo {author} {\bibfnamefont {A.}~\bibnamefont
  {Rasmussen}}, \bibinfo {author} {\bibfnamefont {S.}~\bibnamefont {Ali}},
  \bibinfo {author} {\bibfnamefont {F.}~\bibnamefont {Bertoldo}}, \bibinfo
  {author} {\bibfnamefont {T.}~\bibnamefont {Deilmann}}, \bibinfo {author}
  {\bibfnamefont {N.~R.}\ \bibnamefont {Knøsgaard}}, \bibinfo {author}
  {\bibfnamefont {M.}~\bibnamefont {Kruse}}, \bibinfo {author} {\bibfnamefont
  {A.~H.}\ \bibnamefont {Larsen}}, \bibinfo {author} {\bibfnamefont
  {S.}~\bibnamefont {Manti}}, \bibinfo {author} {\bibfnamefont {T.~G.}\
  \bibnamefont {Pedersen}}, \bibinfo {author} {\bibfnamefont {U.}~\bibnamefont
  {Petralanda}}, \bibinfo {author} {\bibfnamefont {T.}~\bibnamefont {Skovhus}},
  \bibinfo {author} {\bibfnamefont {M.~K.}\ \bibnamefont {Svendsen}}, \bibinfo
  {author} {\bibfnamefont {J.~J.}\ \bibnamefont {Mortensen}}, \bibinfo {author}
  {\bibfnamefont {T.}~\bibnamefont {Olsen}},\ and\ \bibinfo {author}
  {\bibfnamefont {K.~S.}\ \bibnamefont {Thygesen}},\ }\bibfield  {title}
  {\bibinfo {title} {Recent progress of the computational 2d materials database
  (c2db)},\ }\href {https://doi.org/10.1088/2053-1583/ac1059} {\bibfield
  {journal} {\bibinfo  {journal} {2D Materials}\ }\textbf {\bibinfo {volume}
  {8}},\ \bibinfo {pages} {044002} (\bibinfo {year} {2021})}\BibitemShut
  {NoStop}%
\bibitem [{\citenamefont {Jo}\ \emph {et~al.}(2014)\citenamefont {Jo},
  \citenamefont {Ubrig}, \citenamefont {Berger}, \citenamefont {Kuzmenko},\
  and\ \citenamefont {Morpurgo}}]{jo2014}%
  \BibitemOpen
  \bibfield  {author} {\bibinfo {author} {\bibfnamefont {S.}~\bibnamefont
  {Jo}}, \bibinfo {author} {\bibfnamefont {N.}~\bibnamefont {Ubrig}}, \bibinfo
  {author} {\bibfnamefont {H.}~\bibnamefont {Berger}}, \bibinfo {author}
  {\bibfnamefont {A.~B.}\ \bibnamefont {Kuzmenko}},\ and\ \bibinfo {author}
  {\bibfnamefont {A.~F.}\ \bibnamefont {Morpurgo}},\ }\bibfield  {title}
  {\bibinfo {title} {{Mono- and Bilayer WS2 Light-Emitting Transistors}},\
  }\href {https://doi.org/10.1021/nl500171v} {\bibfield  {journal} {\bibinfo
  {journal} {Nano Lett.}\ }\textbf {\bibinfo {volume} {14}},\ \bibinfo {pages}
  {2019} (\bibinfo {year} {2014})}\BibitemShut {NoStop}%
\bibitem [{\citenamefont {Zhang}\ \emph {et~al.}(2015)\citenamefont {Zhang},
  \citenamefont {Chen}, \citenamefont {Johnson}, \citenamefont {Li},
  \citenamefont {Li}, \citenamefont {Mende}, \citenamefont {Feenstra},\ and\
  \citenamefont {Shih}}]{zhang2015}%
  \BibitemOpen
  \bibfield  {author} {\bibinfo {author} {\bibfnamefont {C.}~\bibnamefont
  {Zhang}}, \bibinfo {author} {\bibfnamefont {Y.}~\bibnamefont {Chen}},
  \bibinfo {author} {\bibfnamefont {A.}~\bibnamefont {Johnson}}, \bibinfo
  {author} {\bibfnamefont {M.-Y.}\ \bibnamefont {Li}}, \bibinfo {author}
  {\bibfnamefont {L.-J.}\ \bibnamefont {Li}}, \bibinfo {author} {\bibfnamefont
  {P.~C.}\ \bibnamefont {Mende}}, \bibinfo {author} {\bibfnamefont {R.~M.}\
  \bibnamefont {Feenstra}},\ and\ \bibinfo {author} {\bibfnamefont {C.-K.}\
  \bibnamefont {Shih}},\ }\bibfield  {title} {\bibinfo {title} {{Probing
  Critical Point Energies of Transition Metal Dichalcogenides: Surprising
  Indirect Gap of Single Layer WSe2}},\ }\href
  {https://doi.org/10.1021/acs.nanolett.5b01968} {\bibfield  {journal}
  {\bibinfo  {journal} {Nano Lett.}\ }\textbf {\bibinfo {volume} {15}},\
  \bibinfo {pages} {6494} (\bibinfo {year} {2015})}\BibitemShut {NoStop}%
\bibitem [{\citenamefont {Chiu}\ \emph {et~al.}(2015)\citenamefont {Chiu},
  \citenamefont {Zhang}, \citenamefont {Shiu}, \citenamefont {Chuu},
  \citenamefont {Chen}, \citenamefont {Chang}, \citenamefont {Chen},
  \citenamefont {Chou}, \citenamefont {Shih},\ and\ \citenamefont
  {Li}}]{chiu2015}%
  \BibitemOpen
  \bibfield  {author} {\bibinfo {author} {\bibfnamefont {M.-H.}\ \bibnamefont
  {Chiu}}, \bibinfo {author} {\bibfnamefont {C.}~\bibnamefont {Zhang}},
  \bibinfo {author} {\bibfnamefont {H.-W.}\ \bibnamefont {Shiu}}, \bibinfo
  {author} {\bibfnamefont {C.-P.}\ \bibnamefont {Chuu}}, \bibinfo {author}
  {\bibfnamefont {C.-H.}\ \bibnamefont {Chen}}, \bibinfo {author}
  {\bibfnamefont {C.-Y.~S.}\ \bibnamefont {Chang}}, \bibinfo {author}
  {\bibfnamefont {C.-H.}\ \bibnamefont {Chen}}, \bibinfo {author}
  {\bibfnamefont {M.-Y.}\ \bibnamefont {Chou}}, \bibinfo {author}
  {\bibfnamefont {C.-K.}\ \bibnamefont {Shih}},\ and\ \bibinfo {author}
  {\bibfnamefont {L.-J.}\ \bibnamefont {Li}},\ }\bibfield  {title} {\bibinfo
  {title} {{Determination of band alignment in the single-layer MoS2/WSe2
  heterojunction}},\ }\href {https://doi.org/10.1038/ncomms8666} {\bibfield
  {journal} {\bibinfo  {journal} {Nat. Commun.}\ }\textbf {\bibinfo {volume}
  {6}},\ \bibinfo {pages} {7666} (\bibinfo {year} {2015})}\BibitemShut
  {NoStop}%
\bibitem [{\citenamefont {Geick}\ \emph {et~al.}(1966)\citenamefont {Geick},
  \citenamefont {Perry},\ and\ \citenamefont {Rupprecht}}]{geick1966}%
  \BibitemOpen
  \bibfield  {author} {\bibinfo {author} {\bibfnamefont {R.}~\bibnamefont
  {Geick}}, \bibinfo {author} {\bibfnamefont {C.~H.}\ \bibnamefont {Perry}},\
  and\ \bibinfo {author} {\bibfnamefont {G.}~\bibnamefont {Rupprecht}},\
  }\bibfield  {title} {\bibinfo {title} {Normal modes in hexagonal boron
  nitride},\ }\href {https://doi.org/10.1103/PhysRev.146.543} {\bibfield
  {journal} {\bibinfo  {journal} {Phys. Rev.}\ }\textbf {\bibinfo {volume}
  {146}},\ \bibinfo {pages} {543} (\bibinfo {year} {1966})}\BibitemShut
  {NoStop}%
\bibitem [{\citenamefont {Stafford}\ \emph {et~al.}(1990)\citenamefont
  {Stafford}, \citenamefont {Schmitt-Rink},\ and\ \citenamefont
  {Schaefer}}]{Stafford1990}%
  \BibitemOpen
  \bibfield  {author} {\bibinfo {author} {\bibfnamefont {C.}~\bibnamefont
  {Stafford}}, \bibinfo {author} {\bibfnamefont {S.}~\bibnamefont
  {Schmitt-Rink}},\ and\ \bibinfo {author} {\bibfnamefont {W.}~\bibnamefont
  {Schaefer}},\ }\bibfield  {title} {\bibinfo {title} {Nonlinear optical
  response of two-dimensional magnetoexcitons},\ }\href
  {https://doi.org/10.1103/PhysRevB.41.10000} {\bibfield  {journal} {\bibinfo
  {journal} {Phys. Rev. B}\ }\textbf {\bibinfo {volume} {41}},\ \bibinfo
  {pages} {10000} (\bibinfo {year} {1990})}\BibitemShut {NoStop}%
\bibitem [{\citenamefont {Koch}\ \emph {et~al.}(1999)\citenamefont {Koch},
  \citenamefont {Cundiff}, \citenamefont {Knox}, \citenamefont {Shah},\ and\
  \citenamefont {Stolz}}]{KOCH1999}%
  \BibitemOpen
  \bibfield  {author} {\bibinfo {author} {\bibfnamefont {M.}~\bibnamefont
  {Koch}}, \bibinfo {author} {\bibfnamefont {S.}~\bibnamefont {Cundiff}},
  \bibinfo {author} {\bibfnamefont {W.}~\bibnamefont {Knox}}, \bibinfo {author}
  {\bibfnamefont {J.}~\bibnamefont {Shah}},\ and\ \bibinfo {author}
  {\bibfnamefont {W.}~\bibnamefont {Stolz}},\ }\bibfield  {title} {\bibinfo
  {title} {Magnetoexciton quantum beats: influence of coulomb correlations},\
  }\href {https://doi.org/https://doi.org/10.1016/S0038-1098(99)00243-4}
  {\bibfield  {journal} {\bibinfo  {journal} {Solid State Commun.}\ }\textbf
  {\bibinfo {volume} {111}},\ \bibinfo {pages} {553} (\bibinfo {year}
  {1999})}\BibitemShut {NoStop}%
\bibitem [{\citenamefont {Kyl\"anp\"a\"a}\ and\ \citenamefont
  {Komsa}(2015)}]{kylanpaa2015}%
  \BibitemOpen
  \bibfield  {author} {\bibinfo {author} {\bibfnamefont {I.}~\bibnamefont
  {Kyl\"anp\"a\"a}}\ and\ \bibinfo {author} {\bibfnamefont {H.-P.}\
  \bibnamefont {Komsa}},\ }\bibfield  {title} {\bibinfo {title} {Binding
  energies of exciton complexes in transition metal dichalcogenide monolayers
  and effect of dielectric environment},\ }\href
  {https://doi.org/10.1103/PhysRevB.92.205418} {\bibfield  {journal} {\bibinfo
  {journal} {Phys. Rev. B}\ }\textbf {\bibinfo {volume} {92}},\ \bibinfo
  {pages} {205418} (\bibinfo {year} {2015})}\BibitemShut {NoStop}%
\end{thebibliography}%

\end{document}